\documentclass[11pt, a4paper]{article}
\pdfoutput=1
\usepackage{jheparxiv}
\usepackage[latin1]{inputenc}
\usepackage{amsmath}
\usepackage{amsfonts}
\usepackage{amssymb}
\usepackage{latexsym}
\usepackage{mathrsfs}
\usepackage{graphicx}
\usepackage{color}
\usepackage{twistor}
\usepackage[all]{xy}

\renewcommand{\d}{\mathrm{d}}

\title{Twistor-strings and gravity tree amplitudes}

\author{Tim Adamo}
\author{and Lionel Mason}

\affiliation{The Mathematical Institute \\ University of Oxford \\
	24-29 St.~Giles', Oxford OX1 3LB \\United Kingdom}
	
\emailAdd{adamo@maths.ox.ac.uk}
\emailAdd{lmason@maths.ox.ac.uk}

\abstract{Recently we discussed how Einstein supergravity tree amplitudes might be obtained from the original Witten and Berkovits twistor-string theory when external conformal gravitons are restricted to be Einstein gravitons.  Here we obtain a more systematic understanding of the relationship between conformal and Einstein gravity amplitudes in that twistor-string theory.  We show that although it does not in general yield Einstein amplitudes, we can nevertheless obtain some partial twistor-string interpretation of  the remarkable formulae recently been found by Hodges and generalized to all tree amplitudes by  Cachazo and Skinner.  The Hodges matrix and its higher degree generalizations encode the world sheet correlators of the twistor string. These matrices control both Einstein amplitudes and those of the conformal gravity arising from the Witten and Berkovits twistor-string.  Amplitudes in the latter case arise from products of the diagonal elements of the generalized Hodges matrices and reduced determinants give the former.  The reduced determinants arise if the contractions in the worldsheet correlator are restricted to form connected trees at MHV.  The (generalized) Hodges matrices arise as weighted Laplacian matrices for the graph of possible contractions in the correlators and the reduced determinants of these weighted Laplacian matrices give the sum of of the connected tree contributions by an extension of the Matrix-Tree theorem.  
}

\begin{document}
\maketitle


\section{Introduction}

The twistor-string introduced by Witten \cite{Witten:2003nn} and
realized as a momentum space tree formula \cite{Roiban:2004yf} for
$\cN=4$ super-Yang-Mills has been a remarkable stimulus for
developments in our understanding of gauge theory amplitudes (see
\cite{Adamo:2011pv} for a review biased towards the interests of these
authors and \cite{Brandhuber:2011ke} and other reviews in that volume
for other influences). The question naturally arises as to whether
analogous ideas can be made to work for gravity.  Berkovits and Witten
showed that
twistor-string theory does contain amplitudes consistent with a
non-minimal form of conformal gravity, a
fourth order conformally invariant gravity theory whose action contains the
square of the Weyl tensor \cite{Berkovits:2004jj}.  However, at this point no twistor-string
theory is known that produces just Einstein supergravity, although the
recent formulae of Cachazo and Skinner
\cite{Cachazo:2012kg,Cachazo:2012pz} (see also
\cite{Bullimore:2012cn,He:2012er}) for $\cN=8$ supergravity amplitudes
from rational curves in twistor space are remarkably suggestive of the
existence of such a theory.
 
In \cite{Adamo:2012nn} it was argued, using an observation of Maldacena \cite{Maldacena:2011mk},
that the twistor-string tree formula for conformal supergravity
amplitudes due to Berkovits and Witten \cite{Berkovits:2004jj} could be
used to calculate Einstein tree amplitudes.
The procedure yields an extra factor
of the cosmological constant $\Lambda$ and hence should vanish as
$\Lambda\rightarrow 0$.  However, for a $n$-particle amplitude, the procedure automatically gives a
polynomial of degree-$n$ in $\Lambda$ so that it is straightforward to
divide by $\Lambda$ and take the limit $\Lambda\rightarrow 0$.  In \cite{Adamo:2012nn}, only three-point MHV and anti-MHV amplitudes were checked in momentum
space (found for all $\Lambda$ and checked in the $\Lambda\rightarrow
0$ limit).  In this article we investigate more general formulae that are
yielded in momentum space in the $\Lambda=0$ limit.  

On one hand, we will see that this basic proposal is in general false.  This failure does not arise from any shortcoming in twistor-string theory, but from the fact that the Maldacena argument only applies to \emph{minimal} conformal supergravity (CSG) and not to the non-minimal version thought to be equivalent to the gravity sector of twistor-string theory.\footnote{This was not appreciated in earlier versions of this paper and we thank David Skinner for this observation; see \S\ref{Discussion} for further discussion of this point.}  Minimal CSG has a global $\SU(1,1)$ symmetry acting on the scalars of the theory, so restricting to Einstein scattering states means that only graviton interactions occur in the bulk and the Maldacena argument applies.  The non-minimal CSG of twistor-string theory does not have this $\SU(1,1)$ symmetry, allowing conformal gravitons to couple with scalars in the bulk.  We will give the acronym BW-CSG to this non-minimal CSG, and note that there is no reason for Maldacena's argument to hold for this theory. Nevertheless, we will still be able to obtain the complete Einstein amplitudes in the MHV sector, and ingredients of these amplitudes beyond MHV, by isolating a subset of the contributions from BW-CSG corresponding to minimal CSG.

We first investigate the $\Lambda=0$ contribution to the amplitudes, and see that this is generically non-vanishing for BW-CSG (see \eqref{BWMHV} below). The contribution appropriate to minimal CSG (and hence to pure Einstein
gravity) should vanish at $\Lambda=0$ due to the extra factor of $\Lambda$ in the action arising from the Maldacena argument.  This vanishing can be obtained by modifying the formulae in \cite{Adamo:2012nn} so that only connected
trees\footnote{Rather than ruling out loops, we can require simply
  that the number of disconnected components amongst the contractions
  is minimal; we thank Mat Bullimore for this observation.} are
allowed amongst the contractions in the worldsheet correlator.  Indeed, the answer is generically
non-vanishing at $\Lambda=0$ if loops and more disconnected terms
amongst the contractions in the
correlation functions are allowed.    The role of the connected tree ansatz should be understood as identifying the minimal conformal supergravity contributions to the full BW-CSG.  This only works at MHV and can be understood from the twistor action for conformal gravity \cite{Mason:2005zm}; a derivation of this will be presented elsewhere \cite{Adamo:2013}.

We move on to consider the $O(\Lambda)$
part of the minimal CSG amplitude, which by Maldacena's
argument should yield the Einstein
gravity contribution at $\Lambda=0$.  Working with just 
connected trees, we see that the tree formulae for the MHV
amplitude of \cite{Nguyen:2009jk} can be identified 
with the Feynman diagrams of contractions required for the worldsheet
correlators restricted to being connected trees.  Furthermore, by summing the
diagrams using a weighted extension of the Matrix-Tree theorem, we can obtain a
more fundamental understanding of the origin of Hodges' recent
remarkable MHV formula \cite{Hodges:2012ym}, as well as one of the
reduced determinant factors in the N$^{k}$MHV formula of Cachazo and
Skinner \cite{Cachazo:2012kg, Cachazo:2012pz}.  These Hodges matrices
can be understood as weighted Laplacian matrices for all possible
contractions extended to a permutation invariant framework.  The
reduced determinants that arise in these formulae come from
extensions of the Matrix-Tree theorem.

The conformal gravity twistor-string formulae are very much rooted in
$\cN=4$ supersymmetry and so do not manifest full permutation
symmetry.  However, if they are correct, they must have an emergent
permutation symmetry presumably arising from a $\cN=8$ formulation of
the momentum space analogue of the tree-formula. As already mentioned,
such a formula has independently been found by Cachazo and Skinner
\cite{Cachazo:2012kg,Cachazo:2012pz}.  In this paper we see how the
Hodges formula and certain key ingredients of the Cachazo-Skinner
formula naturally arise from the Berkovits-Witten twistor-string
theory.  We obtain a fairly complete picture at MHV but rather less
complete at higher MHV-degree.

We now give a brief review of the Cachazo-Skinner formula for comparison to the formulae obtained here.  It
is a natural extension of Hodges' formula for the MHV amplitude, and is a function of the kinematic invariants of the momenta
$\lambda_{i}\tilde\lambda_{i}$ where $i=1,\ldots ,n$ index the
external particles and $\lambda_i$ is a two-component Weyl spinor
$\lambda_{iA}$, $A=0,1$.  For $\tilde \lambda_i$, we understand it
to also include the fermionic momenta so that
$\tilde\lambda_i=(\tilde\lambda_{iA'},\eta_i^a)$ where $A'=0'1'$ are
spinor indices and $a=1,\ldots 8$ is a $R$-symmetry index. It also
uses auxiliary variables $\sigma_i=(\sigma_{i0},\sigma_{i1})$ that
are homogeneous coordinates for $n$ marked points on the Riemann
sphere (the string worldsheet).  We use square brackets for primed
spinor contractions, angle brackets for unprimed spinor contractions
and round brackets for contractions of the $\sigma$s.  

Following \cite{Hodges:2012ym,Cachazo:2012kg,Cachazo:2012pz}, we introduce a $n\times n$ matrix function of the kinematic invariants and $\sigma_{i}$s at N$^{k}$MHV degree ($R$-charge $k+2$) by 
\be{GMatrix0}
\tilde{\phi}^{k}_{ij}=\left\{
\begin{array}{c}
\frac{[i j]}{(i\, j)}\left(\frac{(\xi\;j)}{(\xi\;i)}\right)^{k+2} \quad i\neq j \\  
- \sum_{l\neq i}\frac{[i\; l]}{(i\;l)}\left(\frac{(\xi l)}{(\xi i)}\right)^{k+2} \quad i= j   
\end{array}\right. ,
\ee
where $\xi$ is an arbitrary fixed point on $\CP^{1}$.  The off-diagonal entries of $\tilde{\phi}^{k}$ are just the twistor-string propagators associated to contractions among worldsheet vertex operators when $\Lambda=0$. This is related to the matrix $\tilde{\Phi}^{k}$ appearing in \cite{Cachazo:2012kg,Cachazo:2012pz} (and in \cite{Hodges:2012ym} for $k=0$) via conjugation by the matrix $T=\mathrm{diag}\left((\xi\; i)^{k+2}\right)$:
\be{GMatrix}
\tilde{\Phi}^{k}_{ij}=\left\{
\begin{array}{c}
\frac{[i j]}{(i\, j)} \quad i\neq j \\  
- \sum_{l\neq i}\frac{[i\; l]}{(i\;l)}\left(\frac{(\xi l)}{(\xi i)}\right)^{k+2} \quad i= j   
\end{array}\right. ,\qquad \tilde{\phi}^{k}=T^{-1}\tilde{\Phi}^{k}T.
\ee
Hence, reduced determinants of $\tilde{\phi}^{k}$ and $\tilde{\Phi}^{k}$ will be equivalent.\footnote{The fact that \eqref{GMatrix} and hence \eqref{GMatrix0} is independent of the choice $\xi\in\CP^{1}$ follows from the delta-function support of the amplitude; see \cite{Cachazo:2012kg,Cachazo:2012pz} for more details.}

When $k=0$, \eqref{GMatrix} is Hodges' matrix for the MHV formula, which has three-dimensional kernel as a consequence of
momentum conservation.  The general $\tilde{\Phi}^{k}$ has a $k+3$-dimensional kernel
given by the relations
\be{}
\sum_{j=1}^{n} \tilde{\Phi}_{ij}^k \sigma_{j}^{A_{1}}\cdots\sigma_{j}^{A_{k+2}}=0\, ,
\ee
which follow from the support of delta functions in the formula.
An analogue of the Fadeev-Popov procedure leads to an invariant determinant
$\det'(\tilde{\Phi}^{k})$ of
this matrix generalizing that described by Hodges.  This is the main
new ingredient in the tree formula beyond
the Yang-Mills case.  The final formula can be expressed as
\be{F1}
\cM_{n,k}(1,\ldots,n) = \int \frac{\prod_{r=0}^{k+1} \d^{4|8}U_{r}}{\mathrm{vol}\;\GL(2,\C)}\mathrm{det}^{\prime}(\tilde{\Phi}^{k})\mathrm{det}'(\Phi^k)\prod_{i=1}^{n}\D\sigma_{i}\;h_{i}\left(\lambda_{i}\tilde{\lambda}_{i}; Z(\sigma_{i})\right),
\ee
where the $U_{r}$ are parameters for the degree $k+1$ map $Z:\CP^{1}\rightarrow\PT$; $\det' (\Phi^{k})$ and $\det'(\tilde{\Phi}^{k})$ are modified determinants; and the $h_{i}$ are momentum eigenstates. 

It is instructive to compare this to the analogous formula for $\cN=4$ Yang-Mills:
\be{FYM}
\cA_{n,k}(1,\ldots,n) =\int  \frac{\prod_{r=0}^{k+1} \d^{4|4}U_{r}}{\mathrm{vol}\;\GL(2,\C)}  \prod_{i=1}^{n}\frac{\D\sigma_{i}}{(i\;i+1)} a_{i}\left(\lambda_{i}\tilde{\lambda}_{i}; Z(\sigma_{i})\right)  .
\ee
As remarked in \cite{Roiban:2004yf}, after integrating out part of the moduli space, both formulae have as many delta functions as integrals and 
the result reduces to a sum of residues at the
solutions (both real and complex) of the equations implied by delta-functions.
The main difference between \eqref{F1} and \eqref{FYM} is that the Parke-Taylor-like denominator of $\cN=4$ is replaced by the product of determinants $\det'(\tilde{\Phi}^{k}) \det'(\Phi^k)$.

This paper derives Hodges' MHV formula (the $k=0$ version of
\eqref{F1}) from twistor-string theory subject to the connected-tree ansatz, and gives an interpretation of the $\det'(\tilde \Phi^{k})$ factor in \eqref{F1} from twistor-string theory.  Indeed, the individual off-diagonal entries of the matrix are precisely all the possible propagators arising from a certain type of contraction in the worldsheet CFT extended to a permutation invariant framework.  The diagonal entries are then what is required to obtain the weighted Laplacian matrix for the graph of all possible contractions.  In Section \ref{Standard} we find the MHV amplitude of BW-CSG restricted to $\Lambda=0$ Einstein momentum eigenstates is
\be{BWMHV}
\cM^{\mathrm{BW-CSG}}_{n,0}(1,\ldots, n)= \int \frac{\d^{4|4}U_{0}\wedge\d^{4|4}U_{1}}{\mathrm{vol}\;\GL(2,\C)}\left(\prod_{i=1}^{n-2}\tilde{\Phi}^{0}_{ii}\right)\;\prod_{j=1}^{n}h_{j}\;\D\sigma_{j}.
\ee 
By applying the connected tree ansatz in Section \ref{MHV}, we are able to extract the overall coefficient of $\Lambda$ required by the Maldacena argument and show that the $O(\Lambda)$ contribution matches the $k=0$ version of \eqref{F1}. 

The connected tree ansatz is investigated beyond MHV in Section \ref{NMHV}.  We will see that at N$^k$MHV, only $n-2-k$ wordsheet contractions should be allowed and the reduced determinant is constructed precisely as the $n-3-k$ minor that effectively sums the $O(\Lambda)$ contribution from trees with just $n-2-k$ such propagators.  Additionally, we are able to interpret the diagonal entries in the matrix $\Phi^{k}$ appearing in \eqref{F1} in a twistor-string context.  This allows us to conjecture the form of the BW-CSG amplitude for arbitrary MHV degree, again at $\Lambda=0$, 
\be{BWNkMHV}
\cM^{\mathrm{BW-CSG}}_{n,k}(1,\ldots, n)= \int \frac{\prod_{r=0}^{k+1} \d^{4|4}U_{r}}{\mathrm{vol}\;\GL(2,\C)}\left(\prod_{i=1}^{n-k-2}\tilde{\Phi}^{k}_{ii}\right)\left(\prod_{i=n-k-1}^{n}{\Phi}^{k}_{ii}\right)\,\prod_{j=1}^{n}h_{j}\;\D\sigma_{j}.
\ee 
This conjecture relies on lemma \ref{taucont} holding at higher MHV degree, for which there are general arguments.  

The twistor-string ingredients cannot be used to assemble $\det' \Phi $, nor the Vandermonde factors in the definition of $\det' \tilde \Phi$ beyond MHV.  This is related to the fact that our connected tree ansatz no longer extracts the minimal CSG content precisely beyond MHV.  We discuss the distinction between BW-CSG and this minimal framework in more detail in Section \ref{Discussion}. However, it should still be possible to extract $n$-point amplitudes for Einstein scattering states in BW-CSG using the framework outlined in this paper.  Of course, of independent interest, the considerations in this paper are also likely to play a role in any yet to be understood $\cN=8$ twistor-string theory for Einstein gravity.


\section{The Twistor-string Formula for Conformal and Einstein Gravity}

Here we briefly summarize the ingredients used in our twistor-string construction; more details can be found in \cite{Adamo:2012nn}.  Note that we abuse notation slightly, since in this section there is a $\SU(4)_{R}$ $R$-symmetry associated with the twistor-string construction.  We will use similar notation for both this $\cN=4$ and the $\cN=8$ $R$-symmetry used elsewhere.

\subsection{Twistor-strings for conformal supergravity}

Non-projective twistor space is $\T=\C^{4|4}$ and projective twistor
space is $\PT\cong\CP^{3|4}=\T/\{Z\sim \e^\alpha Z\}$, for $\alpha\in\C$.  A twistor will be
represented as $Z^I\in\T$, $Z^I=(Z^\alpha,\chi^a)$, $\alpha=0,\dots 3$,
$a=1,\ldots,4$ with $Z^\alpha$ bosonic and $\chi^a$ fermionic and the
bosonic part $Z^\alpha=(\lambda_{A}, \mu^{A'})$, $A=0,1$, $A'=0'1'$.  A
point $(x,\theta)=(x^{AA'},\theta^{Aa})$ in chiral super Minkowski
space-time $\M$ corresponds to the $\CP^1$ (complex line) $X\subset
\PT$ via the incidence relation \be{incidence}
\mu^{A'}=ix^{AA'}\lambda_{A}\, , \quad \chi^a=\theta^{Aa}\lambda_{A}\, .
\ee with $\lambda_{A}$ homogeneous coordinates along $X$.

We use a closed string version of the Berkovits model\footnote{The standard Berkovits open twistor-string
  would be just as good for most of the considerations in this paper;
  although it is tied into split signature and gets some signs wrong,
  subtleties over choices of contours are avoided. For more on the version used here see \cite{Mason:2007zv}.}
\cite{Berkovits:2004hg,Berkovits:2004jj} with a Euclidean worldsheet $\Sigma$.  The fields are
$$
Z:\Sigma\rightarrow \T\, , \quad Y\in \Omega^{1,0}(\Sigma)\otimes T^*\T\, , \quad \mbox{ and } \quad a\in \Omega^{0,1}(\Sigma)\, .
$$ 
The action is
\be{string-action}
S[Z,Y,a]=\int_\Sigma Y_{I}\dbar Z^{I} + a Z^{I}\; Y_{I} \, ,
\ee
up to matter terms, and has the gauge freedom
$$
(Z,Y,a)\rightarrow (\e^\alpha Z,\e^{-\alpha}Y, a-\dbar \alpha), \qquad \alpha\in\C .
$$
The gauging reduces the string theory to one in $\PT$ and the formalism allows one to use homogeneous coordinates on $\PT$. 

Amplitudes are computed as an integral of worldsheet correlators of vertex operators on $\Sigma$ over the moduli space of `zero-modes': the space of classical solutions to the equations of motion.  For gravity, the vertex operators correspond to deformations of the complex structure together with the cohomology class of deformations of the $B$-field.  These are given by $\dbar$-closed $(0,1)$-forms $F$ on the bosonic part of twistor space $\PT$ with values in the generalized tangent space $ T\oplus T^* \PT$. On $\T$  we represent these by $\dbar$-closed $(0,1)$-forms $F:=(f^I,g_I)$ of homogeneity $(1,-1)$ satisfying $\p_I f^I=0=Z^Ig_I$, defined modulo gauge transformations $(\alpha Z^I,\p_I\beta)$.  These conditions imply that $(f^I\p_I,g_I\d Z^I) $ represents a section of $T\oplus T^*\PT$.  The corresponding vertex operators take the form
$$
V_F:= V_f+V_g:=\int_\Sigma f(Z)^IY_I + g(Z)_I\d Z^I \, .
$$

For $n$-particle tree-level amplitudes, we take $\Sigma\cong\CP^1$ and the amplitude reduces to
an integral of a correlation function of $n$ vertex operators
\be{amplitudes}
\cM(1,\ldots,n)=\sum_{d=0}^\infty \int_{\CM_{d,n}} \d\mu_d \la V_{F_1} \ldots V_{F_n}\ra _d \, ,
\ee
over the space  $\CM_{d,n}$ of maps $Z:\CP^1\rightarrow\PT$ of degree-$d$ and $n$ marked points.\footnote{The rules for taking the correlators are different at different degrees, hence the subscript $d$ on the correlator.}  This is the moduli space of zero-modes for the twistor-string, and has a mathematical definition as a supersymmetric generalization of Kontsevich's moduli space of stable maps \cite{Adamo:2012cd}.  To be more concrete, we can represent the maps by
\be{measure}
Z(\sigma)=\sum_{r=0}^d \sigma_0^r\sigma_1^{d-r} U_r\, , \quad \d \mu_d =\frac{1}{\mathrm{vol}\GL(2,\C)} \prod_{r=0}^d \d^{4|4}U_r \;,
\ee
where $\sigma_{A}$ are homogeneous coordinates on $\CP^1$ and $U_r\in \T$ provide a set of coordinates on $\CM_{d,0}$ with redundancy $\GL(2,\C)$ acting on the $\sigma$ and hence the $U_r$.  The vertex operator $V_{F_i}=V_F(Z(\sigma_i))$ is inserted at the $i$th marked point $\sigma_i\in\Sigma$, and the correlator  naturally introduces a $(1,0)$-form at each marked point either from the $Y_{I}$ or the $\d Z^I$, whereas the `wave-functions' $(f^I,g_I)$ naturally restrict to give a $(0,1)$-form at each marked point.  See \cite{Nair:2007md} for further explanation. 

The correlators are computed by performing Wick contractions of all the $Y$s with $Z$s to give the propagator
\be{OPE}
\la Y(\sigma)_IZ^J(\sigma')\ra_{d}= \left(\frac{(\xi \sigma')}{(\xi \sigma)}\right)^{d+1}\frac{\delta^J_I \D \sigma}{(\sigma \sigma')} \, ,\quad \D\sigma =(\sigma \;\d\sigma)\, , \quad (\sigma\;\sigma')= \sigma_0\sigma_{1}'-\sigma_{1}\sigma_{0}'\, ,
\ee
where $(\sigma \sigma')=\epsilon_{AB}\sigma^{A}\sigma'^{B}$ is the $\SL(2,\C)$ invariant inner product on the world-sheet coordinates.  When $Y$ acts on a function of $Z$ at degree $d$, it then differentiates before applying the contraction; $Y$ acting on the vacuum gives zero so that all available $Y$s must be contracted, but this contraction can occur with any available $Z$.  The $\xi$ is an arbitrary point on the Riemann sphere and reflects the ambiguity in inverting the $\dbar$-operator on functions of weight $d$ on $\Sigma$.  The overall formula should end up being independent of the choice of $\xi$.

Unlike the Yang-Mills case, the degree $d$ of the map is not directly related to the MHV degree of the amplitude (which essentially counts the number of negative helicity gravitons minus 2).  The MHV degree of an amplitude counts the number of insertions of $V_g$ minus 2 (so the MHV amplitude has two $V_g$s).   Conformal supergravity amplitudes have been calculated from this formula in \cite{Ahn:2005es,Dolan:2008gc}.


\subsection{Reduction to Einstein gravity}

Linearized Einstein gravity solves the linearized conformal gravity equations for all values of the cosmological constant $\Lambda$.  
According to  the Maldacena argument \cite{Maldacena:2011mk}, the classical S-matrix for minimal conformal supergravity (i.e., the action evaluated perturbatively on in- and out-wave functions) agrees with that for Einstein gravity when evaluated on Einstein inout states.  In that paper $\Lambda$ was normalised.  If we make it explicit so as to study the $\Lambda\rightarrow 0$ limit,  the minimal conformal supergravity S-matrix yields the Einstein one  multiplied by an additional factor of the cosmological constant $\Lambda$.

Even for Einstein gravity without supersymmetry, we will need to use the supergeometry of $\cN=4$ supertwistor space.\footnote{At tree-level we can pull out the pure gravity parts of the amplitude, but their construction still relies on the fermionic integration
built into the twistor-string formulae; this simply leads to some spinor contractions.}  We first give the restriction required of the vertex operators for $\cN=4$ supersymmetry and then for $\cN=0$.

To reduce to the Einstein case we must break conformal invariance.
This is done by introducing skew infinity twistors $I_{IJ}$, $I^{IJ}$ with super-indices $I,J=(\alpha, a)$.
The bosonic parts $I_{\alpha\beta}$, $I^{\alpha\beta}$ satisfy  
$$
I^{\alpha\beta}=\frac12\varepsilon^{\alpha\beta\gamma\delta}I_{\gamma\delta}\, , \quad I^{\alpha\beta}I_{\beta\gamma}=\Lambda \delta^\alpha_\gamma\; ,
$$ 
where $\Lambda$ is the cosmological constant and we shall set the fermionic part equal to zero.  In terms of the
spinor decomposition of a twistor $Z^\alpha=(\lambda_{A}, \mu^{A'})$ we
have
\be{infinity}
I_{\alpha\beta}=\begin{pmatrix}  \varepsilon^{AB}& 0\\ 0 &
 \Lambda \varepsilon_{A'B'}\end{pmatrix}\, , \qquad
I^{\alpha\beta}=\begin{pmatrix} \Lambda\varepsilon_{AB} & 0\\ 0 & \varepsilon^{A'B'}\end{pmatrix}\:.
\ee
They have rank two when $\Lambda=0$
(i.e., the cosmological constant vanishes) and four otherwise.  The fermionic parts of $I^{IJ}$ can be non-zero and correspond to a gauging for the $R$-symmetry of the supergravity \cite{Wolf:2007tx, Mason:2007ct}.

Geometrically $I^{IJ}$ and $I_{IJ}$ respectively define a Poisson structure $\{,\}$ of weight $-2$ and contact structure $\tau$ of weight $2$ by
$$
\{h_1,h_2\}:=I^{IJ}\p_I h_1\p_J h_2\, , \qquad \tau=I_{IJ}Z^I\d Z^J\, ,
$$ 
and we can use the Poisson structure to define Hamiltonian vector fields $X_h=I^{IJ}\p_Ih\p_J$ which are homogeneous when $h$ has weight 2.  Lines in $\PT$ on which the contact form $\tau$ vanishes correspond to points at infinity.  This zero-set defines a surface $\scri$ in space-time which is null when $\Lambda=0$, space-like for $\Lambda>0$, and time-like for $\Lambda<0$.

The Einstein vertex operators $(V_h,V_{\tilde h})$ correspond to $V_f+V_g$ subject to the restriction $$(f^I,g_I)= (I^{IJ}\p_J h, \tilde{h} I_{IJ}Z^J)$$ so that
\be{Einstein-V}
 V_{h}=\int_\Sigma I^{IJ}Y_I \p_J h\, , \quad V_{\tilde{h}}=\int_\Sigma \tilde{h} \wedge \tau\, .
\ee
The first of these gives the $\cN=4$ multiplet containing the negative helicity graviton and the second that containing the positive helicity graviton. See \cite{Adamo:2012nn} for more discussion.  In order to reduce to standard non-supersymmetric Einstein gravity we must impose 
\be{0susy}
h=e_2\, , \quad \mbox{ and }\quad \tilde{h}=\chi^4 \tilde{e}_{-6}\, .
\ee
Thus, evaluated on vertex operators constructed from \eqref{Einstein-V} with \eqref{0susy}, \eqref{amplitudes} leads to the construction of Einstein gravity tree amplitudes. With this restriction, we see that there is now a correlation between degree of the maps and MHV degree, as fermionic variables only come with $\tilde{h}$: since there are $4d$ fermionic integrations in the path integral for the amplitude, there must be $d$ insertions of $\tilde{h}$.  

At $\cN=4$ there are spurious amplitudes that can be constructed from the Einstein wave functions corresponding to other conformal supergravity sectors.  In \cite{Adamo:2012nn}, we argued that the Einstein gravity amplitudes could be isolated by imposing the correspondence between the degree of the map and the MHV degree: $d=k+1$.  This leads to the following starting point for tree-level scattering amplitudes in Einstein gravity from twistor-string theory:
\begin{eqnarray}
\cM_{n,k}(1,\ldots,n)&=& \lim_{\Lambda\rightarrow 0}\frac{1}{\Lambda}\int_{\CM_{k+1,n}}\d\mu_{k+1}\left\la V_{h_{1}}\cdots V_{h_{n-k-3}} V_{\tilde{h}_{n-k-2}}\cdots V_{\tilde{h}_{n}}\right\ra_{k+1} \nonumber \\
&=&\lim_{\Lambda\rightarrow 0}\frac{1}{\Lambda}\int_{\CM_{k+1,n}}\d\mu_{k+1} \left\la Y\cdot\partial h_{1}\cdots Y\cdot \partial h_{n-k-3}\;\tilde{h}_{n-k-2}\tau_{n-k-2}\cdots \tilde{h}_{n}\tau_{n}\right\ra_{k+1} \nonumber  \\
&=&\lim_{\Lambda\rightarrow 0}\frac{1}{\Lambda}\int_{\CM_{k+1,n}}\d\mu_{k+1}\; \mathcal{C}_{n,k+1}\, , \label{TSform} 
\end{eqnarray}
where $Y\cdot\partial h_{i}=I^{IJ}Y_{I\;i}\partial_{J\;i}h_{i}$ and the last equation defines $\mathcal{C}_{n,k+1}$ as the worldsheet correlation function of the relevant vertex operators. 

In this paper we will focus on the $\Lambda\rightarrow 0$ limit, in particular on the $O(\Lambda^0)$ and $O(\Lambda)$ parts.  However, according to the Maldacena argument, if the Berkovits-Witten formulation correctly gives conformal supergravity amplitudes, then \eqref{TSform} will do so for Einstein gravity for all $\Lambda$.


\section{Evaluating the Twistor-string Tree Formula}
\label{Standard}

We now turn to the evaluation of the formula \eqref{TSform} to test the claim that it will produce Einstein gravity amplitudes correctly. The most elementary consequence is that at $\Lambda=0$, $\mathcal{C}_{n,k+1}$ must vanish because of the overall extra factor of $\Lambda$.  We will also be interested in the coefficient of $\Lambda$ since according to the Maldacena argument, this will give the $\Lambda=0$ Einstein amplitudes which are now well known.  


\subsection{The $O(\Lambda^0)$ contribution in BW-CSG}

At $O(\Lambda^0)$ we exclude any contraction of a $Y_i$ with a $\tau$; this is because any such contraction will produce a factor of $I^{IJ}I_{JK}$, which vanishes for $\Lambda=0$. Hence, a $Y_i$  can only contract with one of the wavefunctions.  In the standard calculation of the worldsheet CFT correlator in the Berkovits-Witten twistor-string, all such contractions are allowed.  However, we will see that the $O(\Lambda^{0})$ portion of this correlator does not vanish even for the MHV amplitude, indicating that it cannot correspond to the minimal CSG required for the Maldacena argument.   

Working at $k=0$ in
\eqref{TSform}, insert the standard momentum eigenstates 
\be{mom-eig}
h_{j}=\int_{\C} \frac{\d s_{j}}{s_{j}^3} \bar{\delta}^{2}(s_{j}
\lambda_{j}-p_{j}) \exp\left(is_{j}
  [[\mu_{j}\tilde{\lambda}_{j}]]\right). 
\ee  
In the $\Lambda\rightarrow 0$ limit we find
\begin{equation*}
\la I^{IJ}Y_{I\;i}\partial_{J}h_{i}\;h_{j}\ra_{1} = \frac{[ij]}{\la i j\ra}\frac{\la \xi j\ra^2}{\la \xi i\ra^2} h_i h_j.
\end{equation*}
After performing all world-sheet
integrals, the na\"{i}ve contribution from the $i$th $Y_{i}$ to the $\Lambda=0$ part of the correlator is:
\begin{equation*}
\sum_{j\neq i}\frac{[ij]}{\la i j\ra}\frac{\la \xi j\ra^2}{\la \xi i\ra^2}\equiv -\phi^{i}_{i},
\end{equation*}
borrowing Hodges' notation \cite{Hodges:2012ym}.  Hodges shows that it follows from momentum conservation that this is independent of $\xi$ and indeed has an interpretation as an inverse soft factor for inserting a particle at the site $i$.  This gives
\begin{equation*}
\mathcal{C}_{n,1}|_{O(\Lambda^0)} =\prod_{i=1}^{n-2}\;-\phi^{i}_{i},
\end{equation*}
but this does not generically vanish, and gives the formula \eqref{BWMHV} from the Introduction.  


\subsection{Connected trees and Feynman diagrams}

Rather than give up and deduce that the Berkovits-Witten twistor-string is incorrect, 
we instead use this as an indication that a different rule is needed for extracting the minimal CSG content.\footnote{In Section \ref{Discussion}, we discuss how this can be viewed in light of the distinction between BW-CSG of the twistor-string and minimal CSG.}  The rule we will impose is that the Feynman
diagrams for the contractions in the correlators appearing in \eqref{TSform} must be
connected trees.  This eliminates the loops and minimises the number of disconnected terms
which are implicitly included in the standard calculation above.  We
will see that it leads to the Hodges \cite{Hodges:2012ym} formula at
MHV via the weighted Matrix-Tree theorem, an analogue of Kirchoff's theorem for
directed weighted graphs (see appendix \ref{graphs} and \cite{Stanley:1999, vanLint:2001, Stanley:2012}).  It
also sheds light on the possible twistor-string origin of Cachazo and
Skinner's recent formula for tree-level N$^k$MHV amplitudes
\cite{Cachazo:2012kg,Cachazo:2012pz}.  As a bonus, it also provides an
understanding of the origins for the tree formulae of \cite{Nguyen:2009jk} and \cite{Bern:1998sv} in twistor-string theory.

Of course, from the point of view of the twistor-string, restricting to this semi-classical connected tree formalism is rather un-natural.  Indeed, after restricting to a genus zero worldsheet, \emph{all} Wick contractions should be allowed in the CFT from the point of view of string theory.  This is a rather puzzling issue which we will not attempt to resolve in this paper; however, there are some hints as to why the connected tree formalism is necessary dealing with the relationship between $\cN=4$ SUGRA and $\cN=4$ conformal supergravity.  We will address this briefly in the discussion of Section \ref{Discussion}.

\begin{figure}[h]
\centering
\includegraphics[width=4.20 in, height=1 in]{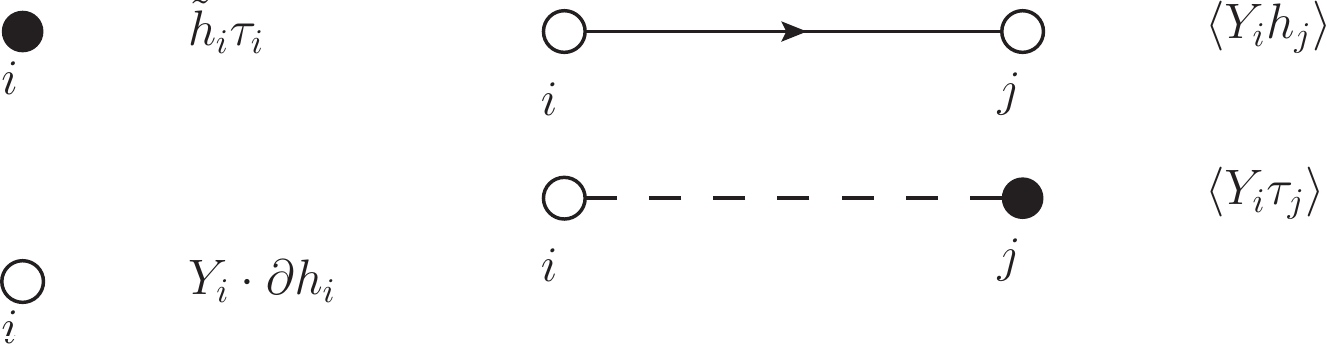}\caption{\small{\textit{Building blocks for Feynman diagrams}}}\label{FeynRules}
\end{figure}

Our argument represents terms contributing to the correlators of
\eqref{TSform} as Feynman diagrams of the worldsheet CFT.  It will be
useful to represent such diagrams graphically, so we set out our
notation here.  There are two different kinds of vertices and two
different kinds of propagator (or contraction) which can contribute to
$\mathcal{C}_{n,k+1}$.  Vertices correspond to either $Y_{i}\cdot \p h_{i}$
(white) or $\tilde{h}_{j}\tau_{j}$ (black).  Straight solid directed
edges correspond to 
propagators arising from contractions directed outwards from a $Y_{i}$
to an Einstein wave function (either $h_{j}$ or $\tilde{h}_{j}$); this utilizes the Poisson structure of the twistor space.
Straight dashed edges correspond to a contraction directed outwards from
a $Y_{i}$ to a  $\tau_j$, and involve the contact structure.  These both arise from standard contractions
in the Berkovits-Witten twistor-string theory \cite{Berkovits:2004jj,
  Nair:2007md}.  See figure \ref{FeynRules}.
We will only allow contributions from diagrams that are maximally connected trees (i.e., no closed cycles). 

The diagrams potentially acquire a factor of $\Lambda^r$ if there are $r$
$\la Y_{i}\tau_{j}\ra$ contractions.  However, it can be seen that such contractions actually vanish at MHV:
\begin{lemma}\label{taucont}
The contractions $\la Y_{i}\tau_{j}\ra_{1}$ vanish and so do not contribute to the correlator $\mathcal{C}_{n,1}$.
\end{lemma}
\proof  The $\la Y_{i}\tau_{j}\ra_{1}$ contractions correspond to $\la \tau (\sigma) V_h\ra_{1}$ at the level of vertex operators. We first note that
\begin{equation*}
\la Z^{I}(\sigma) V_f\ra_{1}=\int_{\CP^1}\frac{\D\sigma'}{(\sigma \sigma')}\frac{(\xi\sigma)^2}{(\xi\sigma')^2}f^{I}(Z(\sigma')).
\end{equation*}
The contraction with $\tau$ is then given by
\begin{multline*}
\la\tau(\sigma) V_{f}\ra_{1} = I_{IJ}\int_{\CP^{1}}\frac{\D\sigma'}{(\sigma\sigma')^2}{(\xi\sigma)}{(\xi\sigma')^2}\left[(\xi\sigma)(\sigma\sigma') \partial Z^{J}(\sigma) f^{I}(\sigma')+ (\xi\sigma)(\sigma' \d\sigma)Z^{I}(\sigma) f^{J}(\sigma') \right. \\
\left. + 2 (\sigma\sigma')(\xi\d\sigma)Z^{I}(\sigma)f^{J}(\sigma')\right] \\
=  I_{IJ}\int_{\CP^{1}}\frac{\D\sigma'}{(\sigma\sigma')^2}{(\xi\sigma)}{(\xi\sigma')^2}\left[(\xi\sigma)(\sigma\sigma') \partial Z^{J}(\sigma) f^{I}(\sigma')+\D\sigma (\xi\sigma') Z^{I}(\sigma)f^{J}(\sigma') \right. \\
\left. +(\sigma\sigma')(\xi\d\sigma) Z^{I}(\sigma)f^{J}(\sigma')\right],
\end{multline*}
with the second expression following by the Schouten identity.

At MHV, the map to twistor space takes the form $Z^{I}(\sigma)=U_{0}^{I}\sigma_{0}+U_{1}^{I}\sigma_{1} = (U \sigma)^{I}$, so $\partial Z^{J}(\sigma)=(U\d\sigma)^{J}$.  Then by the Schouten identity,
\begin{equation*}
\partial Z^{J}(\sigma)\;(\sigma\sigma')=Z^{J}(\sigma')\;\D\sigma - Z^{J}(\sigma)\;(\sigma'\d\sigma),
\end{equation*}
and feeding this into the above expression leaves us with
$$
\la \tau (\sigma) V_f\ra_{1} = I_{IJ} \D\sigma\int_{\CP^1}\frac{\D\sigma'}{(\sigma\sigma')^2}\frac{(\xi\sigma)}{(\xi\sigma')^2}\left(2(\xi\sigma') \, Z^{I }(\sigma)- (\xi\sigma) \,Z^{I}(\sigma')\right)f^{J}(Z(\sigma')).
$$ 
So if $f^I=I^{IJ}\p_J h$ is an Einstein wave function, we obtain a contraction between two infinity twistors which gives $I_{IJ}I^{JK}=\Lambda \delta _I^K$ and we obtain 
\begin{multline*}
\la \tau (\sigma) V_h\ra_{1}=\Lambda\D\sigma \int_{\CP^{1}}\frac{\D\sigma'\;(\xi\sigma)}{(\sigma\sigma')^{2}(\xi\sigma')^{2}}\left(2(\xi\sigma')Z^{I}(\sigma)-(\xi\sigma)Z^{I}(\sigma')\right)\partial_{I} h(\sigma') \\
=2\Lambda \D\sigma \int_{\CP^1}\frac{\D\sigma'\;(\xi\sigma)}{(\sigma\sigma')^{2}(\xi\sigma')^{2}}\left((\xi\sigma')\sigma\cdot\partial' h(\sigma')-(\xi\sigma)h(\sigma')\right) \\
=2\Lambda\D\sigma \sigma_{A}\int_{\CP^1}\frac{\partial}{\partial\sigma'_{A}}\left(\frac{\D\sigma'\;(\xi\sigma)h(\sigma')}{(\sigma\sigma')^{2}(\xi\sigma')}\right) \\
= 2\Lambda \D\sigma \sigma_{A}\int_{\CP^1}\partial' \left(\frac{\sigma^{\prime A}(\xi\sigma)h(\sigma')}{(\sigma\sigma')^{2}(\xi\sigma')}\right)=0.
\end{multline*}
In the second line we have used the homogeneity relation, chain rule, and the linearity of $Z(\sigma')$ in $\sigma'$ to deduce that $\sigma\cdot \p_{\sigma'}h(Z(\sigma'))=Z^I(\sigma)\p_I h(Z(\sigma'))$.  
Thus contractions of $\tau$ with $V_h$ vanish.\footnote{ This calculation can also be understood in terms of building a Picard-iterative solution to the equation
\begin{equation*}
\dbar Z^{I}(\sigma)=I^{IJ}\partial_{J}h(Z(\sigma))
\end{equation*}  For a rational curve of degree 1 with respect to the complex structure deformed by the Hamiltonian vector field of $h$.  The preservation of $\tau$ under such a deformation is equivalent to the vanishing contraction noted above.
Further details will appear in \cite{Adamo:2013}.}     $\Box$  

\medskip

\begin{figure}[t]
\centering
\includegraphics[width=4.20 in, height=2 in]{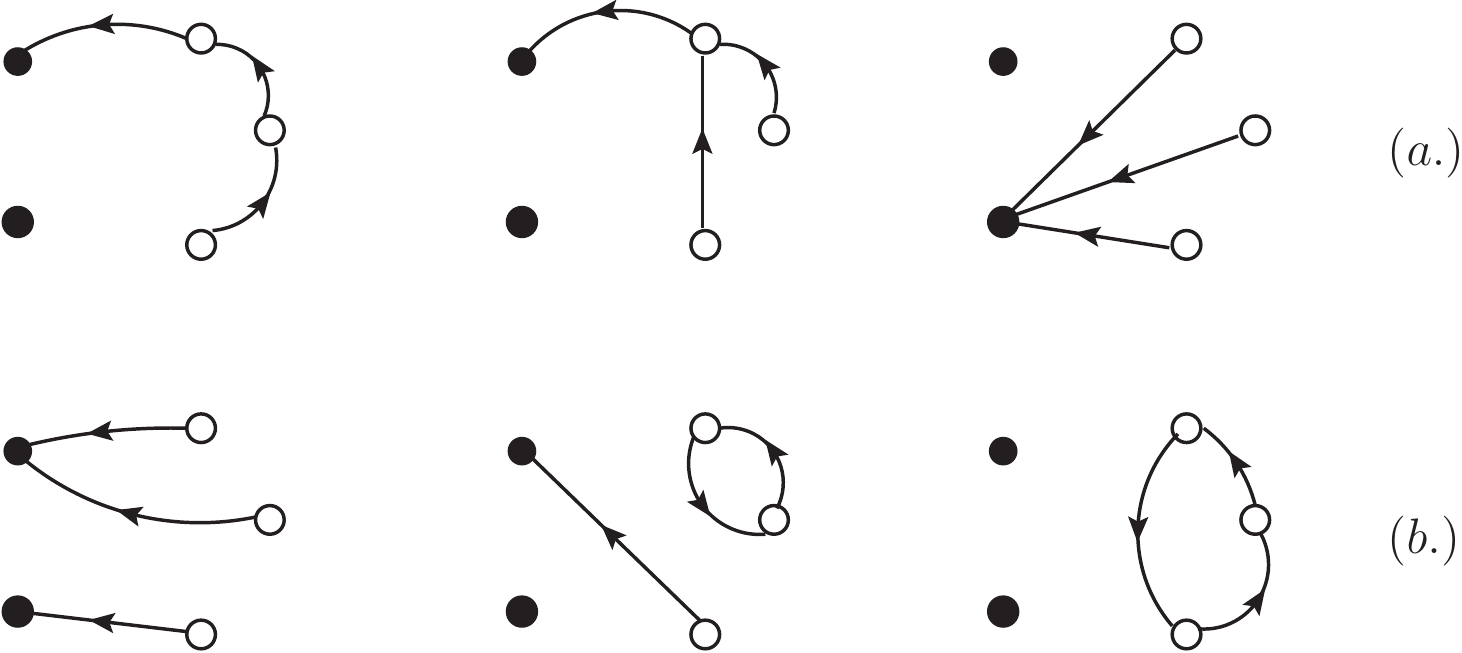}\caption{\small{\textit{Feynman diagrams for the 5-point MHV correlator which are admitted \emph{(a.)}, and excluded \emph{(b.)} from the semi-classical connected tree formalism.}}}\label{Diags}
\end{figure}

This lemma has a geometric interpretation that suggests that it should
generalize to arbitrary MHV degree.  The contraction $\la\tau
V_{f}\ra$ corresponds to the deformation of $\tau$ restricted to a
holomorphic rational curve under the complex deformation of $\PT$
determined by $f\in H^1(\PT,T\PT)$.  In the Einstein case these
deformations preserve the contact structure $\tau$.  This suggests
that lemma \ref{taucont} follows for general reasons and should hold
for all MHV degrees.
 
Hence, we are only left with the Wick contractions between the $Y$s
and the wavefunctions.  So all the Feynman diagrams in our tree
ansatz take the form of directed, connected tree graphs.  Some
examples of graphs which (a)  appear and (b) are excluded by this
ansatz appear in Figure \ref{Diags}. We now investigate these in detail for MHV amplitudes, and generalize to the N$^k$MHV case in section \ref{NMHV}.


\section{The MHV Amplitude}
\label{MHV}

For $k=0$ the diagrams will have 2 black and $n-2$ white vertices.  By lemma \ref{taucont}, there are no $\la Y_{i}\tau_{j}\ra$ contractions and the only edges in play are the directed ones from a vertex operator with a $Y$ to any twistor wave function.  We begin by computing the $\Lambda=0$ portion of the correlator $\mathcal{C}_{n,1}$ in the connected tree formalism, and confirm that it vanishes.  We then derive Hodges' formula for the flat-space MHV amplitude from the $O(\Lambda)$ part of the correlator.


\subsection{The $O(\Lambda^0)$ case}

\begin{figure}
\centering
\includegraphics[width=4 in, height=1 in]{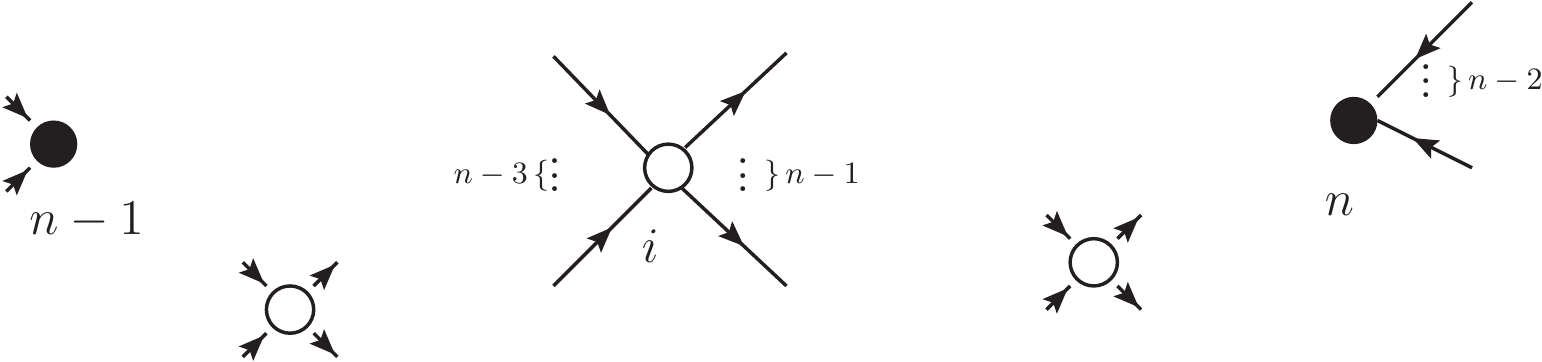}\caption{\small{\textit{The graph G features all possible contractions}}}\label{G}
\end{figure}

We start by drawing a directed graph $G$ which features \emph{all} possible edges at once: each white vertex has $n-1$ outgoing edges (see figure \ref{G}) to each of the other vertices, but just $n-3$ incoming from the remaining white vertices.  Each black vertex has just the incoming $n-2$ edges from each of the white vertices.  The $(n-2)(n-1)$ edges of $G$ can be labelled by a pair $(i,j)$ for the edge starting on white vertex $i$ and ending on $j\neq i$.  

The Matrix-Tree theorem in its simplest form allows us to count the
number of oriented spanning trees of $G$; later we will weight the
counting by the contribution from all the contractions.  The key
device is the $n\times n$ Laplacian matrix of $G$ constructed as follows.

We first define the two modified incidence matrices, $E$ and
$\tilde{E}$ for $G$ with rows indexed by the vertices $v$  of $G$ and columns by the edges $e$.  The  $(v,e)$ entry of $E$ is $1$ if $e$ is outgoing from $v$  and $0$ otherwise, whereas $\tilde{E}$ has $+ 1$ if $e$ is outgoing and $-1$ if $e$ is incoming and $0$ otherwise.  For $G$, the components are given by:
\be{inc1}
E_{i\;(j,k)}=\left\{
\begin{array}{c}
1 \;\mbox{if}\; i=j \\
0 \;\mbox{otherwise}
\end{array}\right. ,\qquad
\tilde{E}_{i\;(j,k)}=\left\{
\begin{array}{c}
1 \;\mbox{if}\; i=j \\
-1 \;\mbox{if}\; i=k \\
0 \; \mbox{otherwise}
\end{array}\right. .
\ee 
The Laplacian matrix is then defined by
\be{mLap1}
L=\tilde{E}E^{\mathrm{T}}=\left(
\begin{array}{cccccc}
n-1 & -1 & \cdots & -1 & 0 & 0\\
-1 & \ddots & & \vdots & \vdots & \vdots \\
\vdots & & & n-1 & 0 & 0 \\
 & & & -1 & 0 & 0 \\
-1 & \cdots & & -1 & 0 & 0 
\end{array}\right).
\ee
The Matrix-Tree theorem (see appendix \ref{graphs}
or \cite{Stanley:1999, vanLint:2001, Stanley:2012} for some
background) enables us to use $L$ to count the Feynman trees
entering into the $O(\Lambda^{0})$ part of the MHV correlator.

\begin{lemma}\label{Counting1}
The number of non-trivial MHV Feynman tree diagrams at $O(\Lambda^0)$ is 
\begin{equation*}
2 \det \tilde{L}_{n-1\;n}=2(n-1)^{n-3},
\end{equation*}
where $\tilde{L}_{n-1\;n}$ is the matrix $L$ with the $n-1^{\mathrm{th}}$ and $n^{\mathrm{th}}$ rows and columns deleted, and $1$ subtracted from the diagonal.
\end{lemma}

\proof The MHV Feynman graphs are spanning trees of the graph $G$
obeying our rules (so that, in particular, there is only one outgoing edge from each white vertex).  By Euler's theorem for a tree the $n-2$ edges connect $n-1$ vertices, so one vertex is left out, and it must be one of the black ones: either 
$n-1$ or $n$.  Consider the class of graphs where $n$ has no incoming edges: delete
the $n^{\mathrm{th}}$ row and column from $L$ and subtract one from the diagonal entries (as each white vertex no longer connects to vertex $n$) to obtain $\tilde{L}_{n}$.  Now,
let $\tilde{L}_{n}$ be the Laplacian matrix of some new graph $G'$.  The
Feynman graphs in this class are the spanning trees of $G'$ rooted at
vertex $n-1$; by the Matrix-Tree theorem (see appendix \ref{graphs}),
these are counted by $\det \tilde{L}_{n-1\;n}$.  By Cayley's theorem, this is equal to $(n-1)^{n-3}$.  The other class is counted in
an equivalent fashion giving the same answer.     $\Box$  

\medskip

To perform the sum of Feynman graphs in the correlator,  we construct a weighted version $\cL$ of $L$ which includes the propagators required by the Feynman rules for the contractions.  We insert a matrix over the set of edges whose diagonal entries are the 
twistor-string propagators at $\Lambda=0$: 
\be{}
 \cP=\mathrm{diag}(\phi^{i}_{j}), \qquad \phi^{i}_{j}=\frac{[ij](\xi\;j)^{2}}{(i j)(\xi i)^{2}},
\ee
so that we can define the weighted Laplacian as
\begin{eqnarray}\label{mLap2}
\cL&=&\tilde{E}\cP E^{\mathrm{T}}, \\
&=&\left(
\begin{array}{cccccc}
\sum_{i\neq 1}\phi^{1}_{i} & -\phi^{1}_{2} & \cdots & -\phi^{1}_{n-2} & 0 & 0\\
-\phi^{2}_{1} & \ddots & & \vdots & \vdots & \vdots \\
\vdots & & & \sum_{i\neq n-2}\phi^{n-2}_{i} & 0 & 0 \\
 & & & -\phi^{n-1}_{n-2} & 0 & 0 \\
-\phi^{n}_{1} & \cdots & & -\phi^{n}_{n-2} & 0 & 0 
\end{array}\right). \notag
\end{eqnarray}

This differs from the matrix $\tilde{\phi}^{0}$ of \eqref{GMatrix0} by a rank 2 error term, and from Hodges' matrix $\tilde{\Phi}$ (the $k=0$ form of \eqref{GMatrix}) by a conjugation and rank 2 error term:
\begin{equation*}
\cL= -\tilde{\phi}^{0}+\mathcal{E}=-T^{-1}\tilde{\Phi}T+\mathcal{E}, \qquad \mathcal{E}=\left(
\begin{array}{ccc}
 & -\phi^{1}_{n-1} & -\phi^{1}_{n} \\
\emptyset & \vdots & \vdots \\
 & \sum_{i\neq n-1}\phi^{i}_{n-1} & -\phi^{n-1}_{n} \\
 & -\phi^{n}_{n-1} & \sum_{i\neq n}\phi^{i}_{n} 
\end{array}\right), \qquad T=\mathrm{diag}\left((\xi\; i)^{2}\right).
\end{equation*}
The matrix $\tilde{\Phi}$ has co-rank 3 thanks to the relation
\be{cr3}
\sum_{j=1}^{n}\tilde{\Phi}_{ij}\sigma_{j}^{A}\sigma_{j}^{B}=0,
\ee
which follows from momentum conservation \cite{Hodges:2012ym}.  Furthermore, the error term is such that once row and columns $n-1$ and $n$ are deleted, the matrices are conjugate.  It is this high co-rank that will enable us to deduce that the sum of diagrams vanishes.

Although Lemma \ref{Counting1} generalizes to tell us how to compute the relevant correlators
by taking determinants of minors of $\cL$, we must divide up into sub-cases according to whether the remaining outgoing propagator from the white vertices lands on the black vertex $n-1$ or $n$.
We first, consider the Feynman trees
rooted at vertex $n-1$; this means that there can be no directed edges to
vertex $n$ and we must delete $\phi^{i}_{n}$ from each diagonal entry.
We therefore define  
\begin{defn} 
Let $\cL^n$ be  $\cL$ with $\phi^{i}_{n}$  deleted from
the $i$th
diagonal entry. We further define $\cL^n_{i\ldots k}$ to be $\cL^n$ with
rows and columns $i, \ldots, k$ deleted.
\end{defn}
For diagrams not involving the vertex $n$, we must also cross out the
$n^{\mathrm{th}}$ row and column of $\cL$ leaving 
the modified $(n-1)\times (n-1)$ matrix denoted $\cL^{n}_{n}$.  The
sum of correlators associated to diagrams in this class is then given
by crossing out the row and column of $\cL^{n}_{n}$ corresponding to
vertex $n-1$ to yield $\cL_{n-1\, n}^n$ and taking its determinant,
thanks to the weighted version of the Matrix-Tree theorem (c.f., appendix \ref{graphs} or \cite{vanLint:2001}, Chapter 36).  A similar
pattern follows for the other class of trees rooted at vertex $n$.
This indicates that the $\Lambda=0$ portion of the conformal gravity
MHV correlator is given by 
\begin{equation*}
\mathcal{C}_{n,1}|_{O(\Lambda^{0})}=\det \cL^{n}_{n\;n-1}+ \det \cL^{n-1}_{n-1\;n}.
\end{equation*}
This sum can be re-interpreted thanks to some basic facts about the matrix $\cL$ and determinants.

\begin{lemma}\label{dets1}
We have 
\be{decomp}
\det \cL^{\ldots i}_{\ldots i\;j}=\sum_{k\neq \ldots i,j}R_{k}\phi^{k}_{j}\, ,\qquad i,j,k=1,\ldots,n\, ,
\ee
where $R_k=\cL^{\ldots i}_{\ldots ijk}$.
\end{lemma}
\proof The matrix $\cL^{\ldots i}_{\ldots ij}$ only depends on the $\phi^{k}_{j}$
through its diagonal entries by the definition of $\cL$ in
\eqref{mLap2}.  We can therefore expand its determinant out along the
diagonal by 
\begin{equation*}
\det \cL^{\ldots i}_{\ldots i\;j}= A+\sum_{k\neq i,j}R_{k}\phi^{k}_{j},
\end{equation*}
where $R_k=\cL^{\ldots i}_{\ldots ijk}$ so it suffices to show that
$A=0$.  By lemma \ref{Counting1} and the Matrix-Tree theorem, we know
that each term in $\det \cL^{\ldots i}_{\ldots i\;j}$ corresponds to the CFT
correlator of a tree rooted at vertex $j$ (omitting vertices $\ldots$
also).  This means that every term
will contain (at least) one propagator to vertex $j$, by definition of
a rooted tree graph.  Thus there can be no term in $\det
\cL^{\ldots i}_{\ldots i\;j}$ of order zero in $\phi^{k}_{j}$ for all $k\neq j$; in other words, $A=0$.     $\Box$

\medskip

\begin{lemma}\label{dets2}
$\det \cL^{n}_{n\;n-1}+ \det \cL^{n-1}_{n-1\;n}=\det \cL_{n-1\;n}$.
\end{lemma}
\proof Lemma \ref{dets1} means that,
\begin{equation*}
\det \cL^{n}_{n\;n-1}= \sum_{i\neq n-1,n}R_{i}\phi^{i}_{n-1}, \qquad \cL^{n-1}_{n-1\;n}= \sum_{i\neq n-1,n}S_{i}\phi^{i}_{n}, 
\end{equation*}
where $R_{i}$ and $S_{i}$ are the remaining factors.  However, the matrices $\cL^{n}_{n\;n-1}$ and $\cL^{n-1}_{n-1\;n}$ differ only by the presence of $\phi^{i}_{n-1}$ and $\phi^{i}_{n}$ along their diagonals, respectively, so it follows that $R_{i}=S_{i}$.  Therefore, we can write
\begin{equation*}
\det \cL^{n}_{n\;n-1}+ \det \cL^{n-1}_{n-1\;n}=\sum_{i\neq  n-1,n} R_{i}(\phi^{i}_{n-1}+\phi^{i}_{n})=\det \cL_{n-1\; n},
\end{equation*}
as required.     $\Box$

\medskip

Lemmas \ref{dets1} and \ref{dets2} therefore give
\be{corrs1}
\det \cL_{n-1\; n}=\det\left(-(T^{-1}\tilde{\Phi}T)_{n-1\;n}+\mathcal{E}_{n-1\;n}\right)=(-1)^{n-2}\det \tilde{\Phi}_{n-1\;n}=0,
\ee
since $\tilde{\Phi}$ has co-rank three and the error term was only
non-vanishing in the $n-1^{\mathrm{st}}$ and $n^{\mathrm{th}}$ columns.  This confirms that
the $\Lambda =0$ contribution to $\mathcal{C}_{n,1}$ vanishes, as
required by the Maldacena relation between conformal and Einstein
gravity amplitudes.  


\subsection{Hodges' formula from the $O(\Lambda)$ contribution}

As $\Lambda\rightarrow 0$, Maldacena's argument also tells us that the flat-space Einstein gravity amplitude should correspond to the $O(\Lambda)$ coefficient of $\mathcal{C}_{n,1}$; this is reflected in \eqref{TSform}.  Hodges' formula for the MHV amplitude in $\cN=8$ supergravity is
\cite{Hodges:2012ym}: 
\be{HF}
\cM^{0}_{n}=\delta^{4|16}\left(\sum_{j=1}^{n}\lambda_{j}\tilde{\lambda}_{j}\right) \mathrm{det}'\left(\tilde{\Phi}^{0}\right),
\ee
where $\mathrm{det}'(\tilde{\Phi}^{0})$ is a reduced determinant of
the MHV ($k=0$) matrix \eqref{GMatrix}, and the delta-function
expresses super-momentum conservation.  Although other forms of the
MHV amplitude have been known for some time (e.g., the BGK or BCFW
type formulas \cite{Berends:1988zp, Mason:2008jy, Bedford:2005yy}),
\eqref{HF} manifests the full permutation symmetry of gravitational
scattering amplitudes without a cumbersome summation, and makes no
reference to any cyclic ordering of the gravitons.  In this sense,
Hodges' formula is the closest gravitational analogue to the
Parke-Taylor formula for gluon scattering in Yang-Mills theory. It was
proved using a $\cN=7$ version of BCFW recursion
\cite{Hodges:2011wm}.  Here we will see how it arises from the twistor-string
formula \eqref{TSform} at $O(\Lambda)$. 

The crucial difference from the $O(\Lambda^{0})$ case is that now the Wick contractions of the twistor-string theory are performed with general infinity twistors:
\begin{equation*}
I^{IJ}\la Y_{I\;i}\partial_{J}h_{i}\; h_{j}\ra_{1} = \frac{I^{IJ}(\xi j)^{2}}{(ij)(\xi i)^{2}}\partial_{I}h_{i}\;\partial_{J}h_{j},
\end{equation*}
with the infinity twistor given by \eqref{infinity} with $\Lambda\neq 0$.  If we continue to use the standard momentum eigenstates \eqref{mom-eig} for the twistor wavefunctions, then these propagators are rather complicated due to derivatives of delta-functions appearing in the $O(\Lambda)$ part of the twistor Poisson structure.  We can avoid this issue if we instead use \emph{dual twistor} wavefunctions of the form:
\be{dtwf}
h_{i}=\int_{\C}\frac{\d t_{i}}{t_{i}^{1+w_{i}}}\exp\left(i t_{i}W_{i}\cdot Z(\sigma_{i})\right), \qquad w_{i}=\left\{
\begin{array}{cc}
-2 & \mbox{if}\;i=n-1, n \\
2 & \mbox{otherwise}
\end{array} \right. .
\ee
Here $W_{i\;I}=(\tilde{\mu}^{A}, \tilde{\lambda}_{i}^{A'})$ are coordinates on $n$ copies of dual twistor space, $\PT^{\vee}$.  These wavefunctions have been used before in other contexts \cite{Mason:2009sa, Cachazo:2012pz}, and can be paired with momentum eigenstates in an appropriate manner to obtain momentum space scattering amplitudes at the end of our calculations.  With the choice of \eqref{dtwf}, the Einstein vertex operators in the twistor-string become:
\begin{eqnarray}
V_{h_{i}} & = & \int\frac{\d t_{i}}{t_{i}^{2}}\;I^{IJ}Y_{i\;I}W_{i\;J}\;\e^{it_{i}W_{i}\cdot Z(\sigma_{i})}, \\
V_{\tilde{h}_{i}} & = & \int t_{i}\;\d t_{i}\; e^{it_{i}W_{i}\cdot Z(\sigma_{i})}\;\tau_{i},
\end{eqnarray}
and the, contractions between the $Y$s and wavefunctions are purely algebraic:
\begin{equation*}
\left\la I^{IJ}Y_{j\;I}W_{j\;J} \e^{it_{i}W_{i}\cdot Z(\sigma_{i})}\right\ra_{1}=it_{i}\frac{I^{IJ}W_{i\;I}W_{j\;J}}{(ij)}\frac{(\xi i)^{2}}{(\xi j)^2} \D\sigma_{j},
\end{equation*}
which limits to the propagator appearing in \eqref{GMatrix0} as $\Lambda\rightarrow 0$:
\be{dtsprop}
\phi^{i}_{j}=i t_{i}t_{j}\frac{I^{IJ}W_{i\;I}W_{j\;J}}{(ij)}\frac{(\xi i)^{2}}{(\xi j)^2}\xrightarrow{\Lambda\rightarrow 0} t_{i}t_{j}\frac{[ij](\xi j)^2}{(ij) (\xi i)^2}.
\ee

By lemma \ref{taucont}, we know that there are no contractions between the $Y$s and $\tau$s, so we can simply write $\tau=I_{IJ}U_{0}^{I}U_{1}^{J}\D\sigma\equiv U^{2}\D\sigma$, and all Wick contractions will involve the propagator \eqref{dtsprop}. This means that we can write \eqref{TSform} as:
\be{DTSform2}
\sum_{\Gamma\subset G}\int\frac{\d^{4|4}U_{0}\wedge \d^{4|4}U_{1}}{\mathrm{vol}\;\GL(2,\C)}(U^{2})^{2}\prod_{j=1}^{n}\frac{\d t_{j}}{t_{j}^{1+w_{j}}}\D\sigma_{j}\; F_{\Gamma}(t_{i}W_{i}, \sigma_{i})\; \e^{i\sum_{j}t_{j}W_{j}\cdot Z(\sigma_{j})},
\ee
where the sum runs over all the Feynman trees $\Gamma\subset G$ contributing to the correlator, and the functions $F_{\Gamma}$ encode the contractions corresponding to each tree diagram.  Lemma \ref{Counting1} tells us precisely how to perform this sum over trees, with the result:
\be{DTSform3}
\int\frac{\d^{4|4}U_{0}\wedge \d^{4|4}U_{1}}{\mathrm{vol}\;\GL(2,\C)}(U^{2})^{2}\prod_{j=1}^{n}\frac{\d t_{j}}{t_{j}^{1+w_{j}}}\D\sigma_{j}\; \mathrm{det}\left(\tilde{\Phi}_{n-1\;n}\right)\; \e^{i\sum_{j}t_{j}W_{j}\cdot Z(\sigma_{j})},
\ee
where the matrix $\tilde{\Phi}$ is now built from the twistor-string propagators of the form \eqref{dtsprop} (which are defined for $\Lambda\neq 0$), and the twice-reduced determinant is non-vanishing because there is no momentum conserving delta function at this stage.

Our task is now to extract the $O(\Lambda)$ portion of this expression.  To do this, we first use the $\GL(2,\C)$ freedom in the measure on $\CM_{1,n}$ to fix the scale and values of $\sigma_{n-1}$ and $\sigma_{n}$ on the $\CP^{1}$ worldsheet.  For the remaining $n-2$ worldsheet coordinates, we can absorb their scalings $t_{i}$ to define non-projective variables $\sigma_{i}t_{i}\rightarrow\sigma_{i}$.  This leaves\footnote{Note that our $\GL(2,\C)$ fixing is implicit in this formulation. An explicit choice such as $\sigma_{n-1}= 0$ and $\sigma_{n}=\infty$ would lead to $Z(\sigma_{n-1})=U_{0}$ and $Z(\sigma_{n})=U_{1}$.}
\begin{equation*}
\int\d^{4|4}U_{0}\wedge \d^{4|4}U_{1} \d^{2}\sigma\;\mathrm{det}\left(\tilde{\Phi}_{n-1\;n}\right)\;(U^{2})^{2} \e^{i\sum_{j}W_{j}\cdot Z(\sigma_{j})},
\end{equation*}  
where we abbreviate
\begin{equation*}
\d^{2}\sigma\equiv \prod_{j=1}^{n-2}\d^{2}\sigma_{j}.
\end{equation*}
The factor of $(U^{2})^2$ makes it awkward to perform the remaining moduli integrals directly, but we can overcome this by representing $U^{2}$ as the operator
\begin{equation*}
\Box=\frac{I_{IJ}}{(n-1\;n)}\frac{\partial}{\partial W_{n-1\;I}}\frac{\partial}{\partial W_{n\;J}} + (n-1\leftrightarrow n),
\end{equation*}
acting on the exponential.  We can then integrate out the moduli to obtain
\be{DTSform4}
\int \;\d^{2}\sigma\;\mathrm{det}\left(\tilde{\Phi}_{n-1\;n}\right)\;\Box^{2}\delta^{8|8}\left(\sum_{i=1}^{n}W_{i}\sigma_{i}\right),
\ee
where the resulting delta function corresponds to momentum conservation.  Note that when $\Lambda=0$, $\Box\rightarrow 1$, so this expression vanishes in accordance with our arguments in the previous subsection.

We know that $\mathrm{det}\left(\tilde{\Phi}_{n-1\;n}\right)$ vanishes to first order on the support of the delta function, so we can integrate by parts to find:
\begin{multline*}
-\int \frac{\d^{2}\sigma}{(n-1\;n)}\frac{\partial}{\partial W_{n\;J}}\mathrm{det}\left(\tilde{\Phi}_{n-1\;n}\right)\;I_{IJ}\frac{\partial}{\partial W_{n-1\;I}}\Box\delta^{8|8}\left(\sum_{i=1}^{n}W_{i}\sigma_{i}\right) + (n-1\leftrightarrow n)\\
=-\Lambda \int \frac{\d^{2}\sigma}{(n-1\;n)}\sum_{i=1}^{n-2}\frac{(\xi n)^{2}}{(in)\;(\xi i)^2} \mathrm{det}\left(\tilde{\Phi}_{n-1\;n\;i}\right)\;W_{i}\cdot\frac{\partial}{\partial W_{n-1}}\Box\delta^{8|8}\left(\sum_{i=1}^{n}W_{i}\sigma_{i}\right) + (n-1\leftrightarrow n).
\end{multline*}
This gives the expected overall factor of $\Lambda$ required, so we can now set $\Lambda=0$ in the remaining portions of this expression.  This means that $\tilde{\Phi}$ agrees with Hodges' matrix (\eqref{GMatrix} for $k=0$), and the remaining $\Box$ operator can be ignored.  

While we have extracted the $O(\Lambda)$ portion of the MHV correlator, our current expression for this is not very illuminating.  Indeed, it appears (at least superficially) to still depend upon the reference spinor $\xi\in\CP^{1}$, which was meant to be arbitrary.  However, further computation shows that this reduces to Hodges' formula (which \emph{is} independent of $\xi$).  

Using the delta function support, we see that $W_{i}\cdot\frac{\partial}{\partial W_{n-1}}$ acts as $\sigma_{n-1}\cdot\frac{\partial}{\partial\sigma_{i}}$.  We can now integrate by parts with respect to $\d^{2}\sigma_{i}$ in each term of the sum to obtain
\begin{multline}\label{DTSform5}
\Lambda \int \frac{\d^{2}\sigma}{(n-1\;n)}\sum_{i=1}^{n-2}\sigma_{n-1}\cdot\frac{\partial}{\partial\sigma_{i}}\left(\frac{(\xi n)^{2}}{(in)\;(\xi i)^2} \mathrm{det}\left(\tilde{\Phi}_{n-1\;n\;i}\right)\right)\;\delta^{8|8}\left(\sum_{i=1}^{n}W_{i}\sigma_{i}\right) + (n-1\leftrightarrow n).
\end{multline}

A straightforward calculation shows that
\begin{multline*}
\sigma_{n-1}\cdot\frac{\partial}{\partial\sigma_{i}}\left(\frac{(\xi n)^{2}}{(in)\;(\xi i)^2}\mathrm{det}\left(\tilde{\Phi}_{n-1\;n\;i}\right)\right) = \\
\left(2\frac{(\xi n)^2 (n-1\xi)}{(in)(\xi i)^3}-\frac{(\xi n)^2 (n-1\;n)}{(in)^{2}(\xi i)^2}\right)\mathrm{det}\left(\tilde{\Phi}_{n-1\;n\;i}\right) \\ 
-(\xi n)^2\sum_{j=1,\;j\neq i}^{n-2}\left(\frac{(n-1\xi)}{(in)(\xi i)(\xi j)^2}+\frac{(n-1\;j)}{(ji)(in)(\xi j)^2}\right)\tilde{\Phi}_{ji}\;\mathrm{det}\left(\tilde{\Phi}_{n-1\;n\;i\;j}\right).
\end{multline*}
We deal with the two sets of terms separately: those with a thrice-reduced determinant, and those with a quadruple-reduced determinant.  The first set of terms is easiest to handle, as two applications of the Schouten identity readily confirm that
\begin{equation*}
\frac{(\xi n)^{2}}{(n\;n-1)(in)(\xi i)^{2}}\left(\frac{(n\;n-1)(\xi i)-2(\xi n-1)(in)}{(in)(\xi i)}\right)+(n-1\leftrightarrow n) =\frac{(n-1\;n)^{2}}{(i n-1)^{2}(n i)^{2}}.
\end{equation*}
This means that the thrice-reduced determinant terms give a contribution to \eqref{DTSform5} of the form
\begin{multline}\label{cont1}
\Lambda \int \d^{2}\sigma \sum_{i=1}^{n-2} \frac{(n-1\;n)^{2}}{(i n-1)^{2}(n i)^{2}} \mathrm{det}\left(\tilde{\Phi}_{n-1\;n\;i}\right)\delta^{8|8}\left(\sum_{i=1}^{n}W_{i}\sigma_{i}\right) \\
= (n-2)\Lambda \int \d^{2}\sigma \frac{(n-1\;n)^{2}}{(i n-1)^{2}(n i)^{2}} \mathrm{det}\left(\tilde{\Phi}_{n-1\;n\;i}\right)\delta^{8|8}\left(\sum_{i=1}^{n}W_{i}\sigma_{i}\right),
\end{multline}
where we have used the defining properties of the reduced determinant $\det'$ of the matrix
$\tilde{\Phi}$ which show that all of these terms will be equal
\cite{Hodges:2012ym}.  This is a consequence of the emergent
permutation symmetry that arises in $\cN=8$ supergravity.

The set of terms appearing with a quadruple-reduced determinant require a more subtle treatment.  Denote their contribution to the correlator as:
\be{4thterms}
-\Lambda\int \d^{2}\sigma\; \sum_{i=1}^{n-2}\sum_{j=1,\;j\neq i}^{n-2}\mathcal{F}^{n-1\;n}_{i\;j}(\xi)\;\delta^{8|8}\left(\sum_{i=1}^{n}W_{i}\sigma_{i}\right)+ (n-1\leftrightarrow n),
\ee
where $\mathcal{F}^{n-1\;n}_{i\;j}(\xi)$ is shorthand for the function
\be{xifunct}
\mathcal{F}^{n-1\;n}_{i\;j}(\xi)=\frac{(\xi n)^{2}}{(n-1\;n)(in)}\left(\frac{(n-1\xi)}{(\xi i)(\xi j)^2}+\frac{(n-1\;j)}{(ji)(\xi j)^2}\right)\tilde{\Phi}_{ji}\;\mathrm{det}\left(\tilde{\Phi}_{n-1\;n\;i\;j}\right).
\ee
While this may seem a complicated mess, the following lemma ensures that we are free to make a judicious choice of $\xi$ which allows us to simplify \eqref{4thterms} considerably:
\begin{lemma}\label{xilemma}
The quantity
\begin{equation*}
\sum_{i=1}^{n-2}\sum_{j=1,\;j\neq i}^{n-2}\mathcal{F}^{n-1\;n}_{i\;j}(\xi)\;\delta^{8|8}\left(\sum_{i=1}^{n}W_{i}\sigma_{i}\right)+ (n-1\leftrightarrow n),
\end{equation*}
is independent of choice of $\xi\in\CP^{1}$.
\end{lemma}
The proof of this fact is a bit technical, so we relegate it to Appendix \ref{Proof}.  

This means that we are free to pick any value for $\xi$ which simplifies \eqref{4thterms}; a particularly good choice is $\xi = \sigma_{n-1}$.  This leads to
\begin{multline}\label{cont2}
\Lambda \int \d^{2}\sigma \sum_{i=1}^{n-2}\sum_{j=1,\;j\neq i}^{n-2}\frac{(n-1\;n)^{2}}{(in)(i\;n-1)(jn)(j\;n-1)}\tilde{\Phi}_{ji}\;\mathrm{det}\left(\tilde{\Phi}_{n-1\;n\;i\;j}\right)\delta^{8|8}\left(\sum_{i=1}^{n}W_{i}\sigma_{i}\right) \\
=(n-2)\Lambda \int \d^{2}\sigma \frac{(n-1\;n)^{2}}{(i n-1)^{2}(n i)^{2}} \mathrm{det}\left(\tilde{\Phi}_{n-1\;n\;i}\right)\delta^{8|8}\left(\sum_{i=1}^{n}W_{i}\sigma_{i}\right),
\end{multline}
where we have once again used the properties of the reduced determinant of $\tilde{\Phi}$ and momentum conservation.  

So we can now combine \eqref{cont1}, \eqref{cont2} to show that (up to an irrelevant numerical constant) the $O(\Lambda)$ portion of the twistor-string correlator is given by:
\be{DTSform6}
\Lambda \int \d^{2}\sigma \frac{(n-1\;n)^{2}}{(i n-1)^{2}(n i)^{2}} \mathrm{det}\left(\tilde{\Phi}_{n-1\;n\;i}\right)\delta^{8|8}\left(\sum_{i=1}^{n}W_{i}\sigma_{i}\right).
\ee
The arguments in the proof of lemma \ref{Counting1} show that this corresponds to summing $(n-2)^{n-4}$ trees of the form appearing in \cite{Bern:1998sv, Nguyen:2009jk}.

At this point, we can easily revert to standard momentum eigenstates (by re-introducing a moduli integral) and insert a power of $(n-1\;n)^{-4}$ by extending the supersymmetry to $\cN=8$
where now this emergent permutation symmetry will extend
over $n$ and $n-1$ also.  This gives the $\cN=8$ momentum-space amplitude:
\begin{multline}\label{MHVamp}
\cM^{0}_{n}= \int_{\CM_{1,n}}\frac{\d^{4|8}U_{0} \wedge\d^{4|8}U_{1}}{\mathrm{vol}\;\GL(2,\C)}\prod_{j=1}^{n}\D\sigma_{j}h_{j}(\lambda_{j}\tilde{\lambda}_{j}; Z(\sigma_{j}))\frac{\det \tilde{\Phi}_{i\;n-1\;n}}{(n\;n-1)^{2}(n-1\;i)^{2}(i\;n)^{2}} \\
=\delta^{4|16}\left(\sum_{j=1}^{n}\lambda_{j}\tilde{\lambda}_{j}\right)\mathrm{det}'(\tilde{\Phi}^{0}).
\end{multline}
Here, the super-momentum-conserving delta function is realized via Nair's lemma \cite{Nair:1988bq}
\begin{equation*}
\delta^{4|16}\left(\lambda\tilde{\lambda}\right)= \int \d^{4|16}x \,\exp\left(i\lambda_{A}\tilde{\lambda}_{A'}x^{AA'}+i\eta_{a}\lambda_{A}\theta^{Aa}\right),
\end{equation*}
and $\det' (\tilde{\Phi}^{0})$ is short-hand for:
\begin{equation*}
\mathrm{det}' (\tilde{\Phi}^{0})\equiv \frac{\det(\tilde{\Phi}_{i\;n-1\;n})}{\la i\;n-1\ra^{2} \la n-1\;n\ra^{2} \la n\;i\ra^{2}},
\end{equation*}
with the $\GL(2,\C)$ freedom fixed by setting $\sigma_{j}=\lambda_{j}$.

Equation \eqref{MHVamp} is just a particular representation of Hodges'
formula \eqref{HF} in the form found by Cachazo and Skinner
\cite{Cachazo:2012kg,Cachazo:2012pz}, so our Feynman tree formalism correctly
produces the MHV amplitude for gravity. 


\section{The N$^{k}$MHV Amplitude}
\label{NMHV}

At N$^k$MHV, the Hodges formula is generalised by the Cachazo-Skinner formula \cite{Cachazo:2012kg,Cachazo:2012pz}.  Like the twistor-string formula, it is an integral over the moduli of rational curves of degree $d=k-1$.  It provides all tree-level N$^{k}$MHV amplitudes of $\cN=8$ supergravity \cite{Cachazo:2012pz} and is given by:  
\be{CSF}
\cM_{n,k}(1,\ldots,n) = \int \frac{\prod_{r=0}^{k+1} \d^{4|8}U_{r}}{\mathrm{vol}\;\GL(2,\C)}\mathrm{det}^{\prime}(\tilde{\Phi}^{k})\mathrm{det}'(\Phi^k)\prod_{i=1}^{n}\D\sigma_{i}\;h_{i}\left(\lambda_{i}\tilde{\lambda}_{i}; Z(\sigma_{i})\right)\, .
\ee
Here, $\tilde{\Phi}^{k}$ is the co-rank $k+3$ matrix \eqref{GMatrix}, and $\Phi^{k}$ is essentially its parity conjugate:
\be{GMatrix2}
\Phi_{ij}=\left\{
\begin{array}{c}
\frac{\la i j\ra}{(i\, j)} \qquad \qquad \qquad \qquad \qquad \qquad \qquad 
i\neq j \\  
- \sum_{l\neq i}\frac{\la i\; l\ra}{(i\;l)}\left(\frac{(\xi l)}{(\xi i)}\right)^{n-k-2}\frac{\prod_{m\neq i}(i\;m)}{\prod_{s\neq l}(l\;s)} \quad i= j   
\end{array}\right. ,
\ee
The $\mathrm{det}^{\prime}(\tilde{\Phi}^{k})$ and $\mathrm{det}'(\Phi^k)$ are modified reduced determinants, involving some additional Vandermonde determinant factors. In particular, $\mathrm{det}^{\prime}(\tilde{\Phi}^{k})$ can be written
\be{VMdet}
\mathrm{det}^{\prime}(\tilde{\Phi}^{k})=\frac{\det\left(\tilde{\Phi}^{k}_{i\;n-k-1\cdots n}\right)}{\prod _{j<k,\;j,k\in\{i,n-k-1,\ldots, n\}}(j\;k)^{2}}\equiv \frac{\det\left(\tilde{\Phi}^{k}_{i\;n-k-1\cdots n}\right)}{\mathrm{VM}(i,n-k-1,\ldots, n)^{2}}.
\ee

That \eqref{CSF} involves an integral over the moduli space of
$n$-pointed stable maps of $\CP^1$ to twistor space is of course highly
suggestive of a twistor-string origin.  We will begin by explaining the factor $\det\tilde{\Phi}^{k}_{i\;n-k-1\cdots n}$ in terms of our connected tree formalism by generalizing arguments from the MHV setting.  The remaining factors should also presumably arise from the other ingredients in the twistor string calculations; we investigate some of these factors later in this section.
 
For the N$^k$MHV amplitude the twistor-string tree formula \eqref{TSform}
\begin{equation*}
\lim_{\Lambda\rightarrow
  0}\frac{1}{\Lambda}\int_{\CM_{k+1,n}}\d\mu_{k+1}\;
\mathcal{C}_{n,k+1}. 
\end{equation*}
is now an integral over the moduli of rational maps of degree $d=k+1$ and there
are now $n-k-2$ vertex operators of the form $Y\cdot \p h$ and $k+2$ of
the form $\tilde h \tau$.
Again we will allow only maximally connected Feynman tree diagrams to contribute to the CFT correlator
$\mathcal{C}_{n,k+1}$; if we do not, then we will again see that we obtain incorrect non-vanishing answers at $\Lambda=0$.  The diagrams therefore now have $k+2$ black vertices and $n-k-2$
white vertices.  In particular there are just $n-k-2$ $Y_i$s to be
contracted with the $h_j$, $\tilde h_j$ or $\tau_j$ and so the
diagrams become correspondingly simpler with fewer edges. The number of edges corresponds to the size of the minors whose determinant we must take when we use the Matrix-Tree theorem.  It turns out that the ranks of the matrices $\tilde{\Phi}^{k}$ given in \eqref{GMatrix0} and \eqref{GMatrix} encoding the twistor-string propagators are $n-k-3$, decreasing as appropriate with $k$.

Again at $\Lambda=0$ we must have vanishing contribution, and for the $O(\Lambda)$ contribution should give the Einstein amplitude.
As before, we draw the master graph $G$ of all possible $Y_i$
contractions with the $h_j$ and $\tilde h_j$.
Thus in $G$, black vertices have $n-k-2$
 incoming edges (one from each of the white vertices) and no outgoing edges, while white vertices will have
 $n-k-3$ incoming (one from each of the other white vertices) and
 $n-1$ outgoing edges to all other vertices.  The definition of the
 modified incidence matrices $E$, $\tilde{E}$ and
 $L=\tilde{E}E^{\mathrm{T}}$ for this new graph follows in precisely
 the same way 
 as in the MHV setting. 

The incidence Laplace matrix of $G$ is now: 
\begin{equation*}
L=\left(
\begin{array}{ccccccc}
n-1 & -1 & \cdots & -1 & 0 & \cdots & 0\\
-1 & \ddots & & \vdots & \vdots & & \vdots \\
\vdots & & & n-1 & 0 & \cdots & 0 \\
 & & & -1 & 0 & \cdots & 0 \\
 & & & \vdots & \vdots & \vdots \\
-1 & \cdots & & -1 & 0 & \cdots & 0 
\end{array}\right),
\end{equation*}
with the last $k+2$ columns composed entirely of zeros.
For the propagators to form a connected tree, there can only be one
outgoing propagator from the white vertices.  If this lands on vertex
$n-k-1$, the other black vertices are isolated and the Matrix-Tree theorem gives the number of such Feynman trees 
rooted at the black vertex $n-k-1$ to be
$\det \tilde{L}_{n-k-1\cdots  n}=(n-k-1)^{n-k-3}$ .

To compute the actual correlator at $O(\Lambda^0)$ rather than just the number of
contributing trees, we define the weighted Laplacian $\cL$ weighted by the
appropriate propagators at degree $d=k+1$ which gives
\begin{equation*}
\cL=\left(
\begin{array}{ccccccc}
\sum_{i\neq 1}\phi^{1}_{i} & -\phi^{1}_{2} & \cdots & -\phi^{1}_{n-k-2} & 0 & \cdots & 0\\
-\phi^{2}_{1} & \ddots & & \vdots & \vdots & & \vdots \\
\vdots & & & \sum_{i\neq n-k-2}\phi^{n-k-2}_{i} & 0 & \cdots & 0 \\
 & & & -\phi^{n-k-1}_{n-k-2} & \vdots & & \vdots \\
 & & & \vdots & & & \\
-\phi^{n}_{1} & \cdots & & -\phi^{n}_{n-k-2} & 0 & \cdots & 0 
\end{array}\right), \qquad \phi^{i}_{j}=\frac{[i\;j]}{(i\; j)}\left(\frac{(\xi\;j)}{(\xi\;i)}\right)^{k+2}.
\end{equation*}
This $\cL$ is related to $\tilde{\Phi}^{k}$ by 
a conjugation by $T=\mathrm{diag}\left((\xi\;i)^{k+2}\right)$ and a
rank $k+2$ error term that disappears when the last $n-k-2$ rows and
columns are deleted.  The natural generalizations of lemmas
\ref{dets1} and \ref{dets2} then tell us that the sum of the Feynman
trees contributing to the correlator $\mathcal{C}_{n,k+1}$ is given
by: 
\begin{equation*}
\det \cL_{n-k-1\cdots n}=\det\tilde{\Phi}^{k}_{n-k-1\cdots n}=0,
\end{equation*}
and this is vanishing because $\tilde{\Phi}^{k}$ has co-rank $k+3$. This gives 
the expected vanishing at $\Lambda=0$.

The Einstein gravity contribution should be given by the
$O(\Lambda)$ contribution, as in the MHV case.  Instead of the MHV pre-factor, we expect to obtain something
more complicated which encodes the data of the Wick contractions, the  $\tau$s and
other Jacobian factors\footnote{In the $k=0$, or $d=1$ case, we were
  able to fix the $\GL(2,\C)$ freedom so that
  $\sigma_{i}=\lambda_{i}$; thereby setting the Jacobian factor equal
  to unity.  For $d>1$ this is no longer possible and a non-trivial
  factors can occur.}.  We will ignore such additional factors here and focus
only from those that arise from contractions of the remaining
$Y$s with the $h$s and $\tilde h$s.  

As in the MHV amplitude, we begin by summing all Feynman trees rooted at any of the $k+2$ $\tau$-vertices, but with $\Lambda\neq 0$.  Performing computations along the lines of those from the $k=1$ setting will eventually produce a sum over $(n-k-2)^{n-k-4}$ modified
trees, which corresponds to taking reduced determinants of the matrix $\cL$
via the natural extensions of lemmas \ref{Counting1}-\ref{dets2}.
Hence, we obtain the contribution $\det\tilde{\Phi}^{k}_{i\;n-k-1\cdots n}$, where $i\in\{1,\ldots,n-k-2\}$.  Of course, we must still sum
over all of the different ways of taking such contributions, as well as accounting for additional factors arising from the various integrations by parts which occur.  As in the MHV case, the properties
of the $\tilde{\Phi}^{k}$ matrix ensure that all of these
contributions are equal up to the kind of Vandermonde factors that we
are ignoring.  For reference, we demonstrate the reduction of the Cachazo-Skinner formula to $\cN=4$ supergravity in appendix \ref{N=4}.

While our connected tree formalism fails to shed light on the reduced determinant $\mathrm{det}'(\Phi^k)$, we conclude this section by noting that the diagonal entries of the matrix $\Phi^{k}$ do have a very natural interpretation in twistor-string theory:
\begin{lemma}
Let $Z:\CP^{1}\rightarrow\PT$ be a map from the abstract worldsheet to twistor space of degree $k+1$, and $\Phi^{k}$ be the matrix given by \eqref{GMatrix2}.  Then the contact structure $\tau(\sigma_{i})\equiv\tau_i$ can be pulled back to the worldsheet:
\begin{equation*}
Z^{*}\;\tau_{i}=-\Phi_{ii}\;\D\sigma_{i}.
\end{equation*}
\end{lemma}

\proof  Since the contour is homologically trivial, we have the following relation between residues:
\begin{equation*}
\sum_{i=n-k-1}^{n}\oint_{|(li)|=\varepsilon}\D\sigma\frac{Z^{I}(\sigma_{i})}{(i\sigma)\prod_{l=n-k-1}^{n}(li)}=\sum_{i=n-k-1}^{n}\frac{Z^{I}(\sigma_{i})}{(i\sigma)\prod_{l\neq i}(li)} =-\frac{Z^{I}(\sigma)}{\prod_{l=n-k-1}^{n}(l\sigma)}.
\end{equation*}
This means that we can write the holomorphic map to twistor space as
\begin{equation*}
Z^{I}(\sigma)=-\sum_{i=n-k-1}^{n}Z^{I}(\sigma_{i})\prod_{l=n-k-1,l\neq i}^{n}\frac{(l\sigma)}{(li)},
\end{equation*}
and hence the contact structure becomes
\begin{multline*}
Z^{*}\tau = I_{IJ}Z^{I}(\sigma)\partial Z^{J}(\sigma)=I_{IJ}\sum_{i,j=n-k-1}^{n}Z^{I}(\sigma_{i})Z^{J}(\sigma_{j})\prod_{l=n-k-1,l\neq i}^{n}\frac{(l\sigma)}{(li)}\partial\left(\prod_{m\neq j}\frac{(m\sigma)}{(mj)}\right) \\
=I_{IJ}\sum_{i,j=n-k-1}^{n}\frac{Z^{I}(\sigma_{i}) Z^{J}(\sigma_{j})}{(ij)}\prod_{l=n-k-1,l\neq i,j}^{n}\frac{(l\sigma)^2}{(li)(lj)}\D\sigma .
\end{multline*}

Now, the matrix $\Phi^{k}$ is not only independent of $\xi\in\CP^{1}$, but can also have $n-k-2$ independent choices of this reference spinor \cite{Cachazo:2012kg, Cachazo:2012pz}.  In particular, this means that the diagonal elements can be written:
\begin{equation*}
\Phi_{ii}=-I_{IJ}\sum_{j\neq i}\frac{Z^{I}(\sigma_{i})Z^{J}(\sigma_{j})}{(ij)}\frac{\prod_{m\neq i}(im)}{\prod_{l\neq j}(jl)}\prod_{r=1}^{n-k-2}\frac{(jp_{r})}{(ip_{r})},
\end{equation*}
where $\{p_{r}\}$ are $n-k-2$ reference points on $\CP^{1}$.  If we choose $\{p_{r}\}_{r=1,\ldots,n-k-2}$ to be equal to $\{\sigma_{i}\}_{i=1,\ldots,n-k-2}$, then this expression is considerably simplified, leaving
\begin{equation*}
\Phi_{ii}=-I_{IJ}\sum_{j=n-k-1,j\neq i}^{n}\frac{Z^{I}(\sigma_{i})Z^{J}(\sigma_{j})}{(ij)}\prod_{l=n-k-1,l\neq i,j}^{n}\frac{(li)}{(lj)}.
\end{equation*}
Then we immediately see
\begin{equation*}
Z^{*}\tau|_{\sigma=\sigma_i}=\tau_{i}=\D\sigma_{i}\;I_{IJ} \sum_{j=n-k-1,j\neq i}^{n}\frac{Z^{I}(\sigma_{i})Z^{J}(\sigma_{j})}{(ij)}\prod_{l=n-k-1,l\neq i,j}^{n}\frac{(li)}{(lj)} = -\Phi_{ii}\;\D\sigma_{i},
\end{equation*}
as required.      $\Box$

\medskip

If we assume that lemma \ref{taucont} holds for maps of degree $k+1$ (i.e., $\la\tau V_{f}\ra_{k+1}=0$), then this immediately provides us with an expression for the N$^k$MHV amplitude of the non-minimal BW-CSG of twistor-string theory.  In particular, if we insert Einstein vertex operators, allow all non-vanishing contractions in the worldsheet CFT, and set $\Lambda=0$, then we are left with \eqref{BWNkMHV}
\begin{equation*}
\cM^{\mathrm{BW-CSG}}_{n,k}(1,\ldots, n)= \int \frac{\prod_{r=0}^{k+1} \d^{4|4}U_{r}}{\mathrm{vol}\;\GL(2,\C)}\left(\prod_{i=1}^{n-k-2}\tilde{\Phi}^{k}_{ii}\right)\left(\prod_{i=n-k-1}^{n}{\Phi}^{k}_{ii}\right)\,\prod_{j=1}^{n}h_{j}\;\D\sigma_{j}.
\end{equation*}
This expression (which is generically non-vanishing) illustrates the distinction between minimal and non-minimal conformal supergravity for generic scattering amplitudes, which is the main obstruction to extracting the remaining ingredients of the Cachazo-Skinner formula using the connected tree formalism.


\section{Discussion}
\label{Discussion}

We have made significant progress in our understanding of the Berkovits-Witten twistor-string evaluated on the Einstein sub-sector of conformal gravity amplitudes and its comparison with the corresponding Cachazo-Skinner formulae for Einstein gravity \eqref{CSF}.
The  moduli space integral and the external wave functions are the same, and indeed the Hodges matrices and their higher degree generalisations and conjugates are useful for expressing both sets of formulae.  However,  in the Berkovits-Witten twistor-string products of the diagonal entries enter, whereas in the Cahachazo-Skinner formulae we use the reduced determinant of $\tilde{\Phi}^{k}$ in the formula. In the case of the generalised Hodges matrices themselves, this could be understood by
requiring that only connected tree diagrams should be taken to contribute to the CFT
worldsheet correlators arising in the twistor-string amplitude formula; the reduced determinant arose by expressing the sum of the resulting Feynman diagrams as the determinant of a weighted Laplacian matrix for the graph of all possible contractions using the weighted Matrix-Tree theorem.  This argument doesn't work beyond MHV as it is not possible to understand the $\tau$ vertex operators as arising from some form of Feynman diagrams and so there is no analogous motivation for the replacement of the product of the diagonal entries by a reduced determinant.  Thus our tree ansatze gives a twistor-string derivation of Hodges' formula, but does not reproduce any of $\det'\Phi$ nor the Vandermonde determinants in the N$^k$MHV  Cachazo-Skinner formula.

  From the string theory perspective it is in any case difficult to motivate the tree ansatze as there are no parameters that can be used to suppress loop contributions at the level of the worldsheet CFT.  However, the geometric interpretation of the restriction to trees is clear: the $Y$ contractions are performing a perturbation expansion for the construction of holomorphic curves in a deformed twistor space given as solutions to
\be{GCE}
\dbar_\sigma  Z^I(\sigma) =I^{IJ}\p_J h(Z)\, .
\ee
The Feynman trees build up the correct classical solution to this equation as appropriate to the Penrose non-linear graviton construction; details of this picture will appear in \cite{Adamo:2013}.  This can be contrasted with the Yang-Mills case, where an exponentiation arguement could be used \cite{Boels:2006ir}.  There the twistor-string na\"ively gives the determinant of the twistor $\dbar$-operator whereas the correct contribution to obtain Yang-Mills is the log of the determinant.

Another perspective on this issue is provided by considering the $\cN=4$ supersymmetry in play at the level of the twistor-string.  The correspondence between Einstein and conformal gravity is stated for $\cN=0$; so long as the conformal supergravity (CSG) we consider for $\cN>0$ is \emph{minimal}, the correspondence should still hold precisely.  Minimal $\cN=4$ CSG possesses a global $\SU(1,1)$ symmetry acting on the scalars of the theory \cite{Bergshoeff:1980is}; this excludes any coupling between the scalars and the Weyl curvature in the Lagrangian (c.f., \cite{Fradkin:1985am}), so restricting to Einstein spin-2 scattering states means that only graviton interactions occur in the bulk.  The perturbative solution to \eqref{GCE} built by the connected tree formalism can be shown to correspond to this minimal CSG in the MHV sector (details will appear in \cite{Adamo:2013}).  Thus the transition to the answers obtained from minimal CSG to BW-CSG seem to correspond to quantising on the world sheet,  

However, there is a conjectured \emph{non-minimal} version of $\cN=4$ CSG which does not posses this $\SU(1,1)$ symmetry \cite{Fradkin:1983tg, Fradkin:1985am}.  This allows for interactions between the scalars and Weyl curvature in the space-time Lagrangian of the form $ c W^2$, leading to three-point vertices coupling two conformal gravitons to a scalar.  In this theory, inserting Einstein scattering states will produce Feynman diagrams involving these vertices which do not correspond to anything in $\cN=4$ supergravity.  While there is some doubt over whether non-minimal conformal supergravity can actually exist in its own right (c.f., \cite{Buchbinder:2012uh}, footnote 4), Berkovits and Witten indicated that twistor-string theory should correspond to non-minimal $\cN=4$ CSG coupled to $\cN=4$ SYM \cite{Berkovits:2004jj}.  Indeed, spurious amplitudes related to the non-minimal coupling between conformal gravitons and scalars were found explicitly in \cite{Adamo:2012nn}.  

Thus, the BW-CSG of the twistor-string is non-minimal and certainly requires modification in order to extract the minimal CSG content and hence the Einstein amplitudes.  This is captured by our expression for the MHV amplitude in BW-CSG \eqref{BWMHV}, which does not vanish at $\Lambda=0$.  The loop and disconnected contributions to the worldsheet correlator (which are dropped in the connected tree formalism) should therefore correspond to space-time Feynman diagrams involving the non-minimal three-point vertices.  Of course, substantial work is required to make these observations precise, since there are many other \emph{a priori} feasible interpretations for these additional contributions.

Finally, we have presented this work as a verification of the Berkovits-Witten twistor-string, but it is also the case that a key motivation is to elucidate the twistor-string underpinnings of the Cachazo-Skinner formula.  Although it is very
suggestive of a twistor-string theory that directly describes $\cN=8$
supergravity, the details have yet to emerge.

\acknowledgments

We would like to thank Mat Bullimore, Freddy Cachazo, Song He,
Andrew Hodges, Parameswaran Nair and especially David Skinner for useful conversations and
comments.   While completing this project, we heard about related work in
progress by Bo Feng and Song He \cite{Feng:2012sy} and Clifford Cheung
\cite{Cheung:2012jz} relating the MHV
tree formula of 
\cite{Nguyen:2009jk} to Hodges' formula.  
TA is supported by a National Science Foundation (USA) Graduate Research Fellowship and by Balliol College; LM is supported by a Leverhulme Fellowship and benefited from hospitality in the Perimeter Institute during some of the work on this project.


\appendix

\section{Graph Theory}
\label{graphs}

In this appendix, we review some graph theory that we have used to sum the Feynman tree graphs.  For a more
complete introduction see e.g., \cite{Stanley:1999, vanLint:2001,
  Stanley:2012}.   

Let $G$ be a directed graph, with a set of vertices $\mathcal{V}$ and directed edges $\mathcal{E}$.  We will denote such a graph by $G=(\mathcal{V},\mathcal{E})$.  Denote the number of vertices as $|\mathcal{V}|=n$ and edges as $|\mathcal{E}|=m$.  The \emph{incidence matrix} of $G$ is a $n\times m$ matrix $E$ encoding the connective structure of the graph.  We can denote an edge of $G$ as $(j,k)\in\mathcal{E}$, meaning the directed edge which is outgoing from vertex $j\in\mathcal{V}$ and incoming to vertex $k\in\mathcal{V}$.  In this notation, the entries of the incidence matrix are given by:
\be{incidencematrix}
E_{i\;(j,k)}=\left\{
\begin{array}{c}
1 \;\mathrm{if}\;i=j \\
-1 \;\mathrm{if}\;i=k \\
0\;\mathrm{otherwise}
\end{array}\right. .
\ee

The incidence matrix can be used to build the \emph{Laplacian matrix} of the graph $G$, which is a $n\times n$ matrix given by:
\be{laplacian}
\Delta =EE^{\mathrm{T}}, \qquad \Delta_{i\;j}=\left\{
\begin{array}{c}
-1 \;\mathrm{if}\;i\neq j\;\mathrm{and}\;(i,j)\in\mathcal{E}\\
d^{\mathrm{out}}_{i}\;\mathrm{if}\;i=j \\
0 \;\mathrm{otherwise}
\end{array}\right. ,
\ee
where $d_{i}^{\mathrm{out}}$ is the out-degree of vertex $i\in\mathcal{V}$.  

The Feynman tree diagrams we are interested in counting throughout this paper are examples of particular sub-graphs of a directed graph, namely rooted spanning trees:
\begin{defn}
A \emph{spanning tree of} $G$ \emph{rooted at} $v\in\mathcal{V}$ is a spanning sub-graph $T=(\mathcal{V},\mathcal{F})$ of $G$ such that: (1.) T has no oriented cycles; (2.) vertex $v$ has out-degree zero; and (3.) every other vertex $i\neq v$ has out degree one.
\end{defn} 

The main result in graph theory we use is the weighted  generalisation of the Matrix-Tree theorem.  The original matrix-tree theorem tells us how to count spanning trees of $G$ rooted at a particular vertex in terms of determinants of the reduced Laplacian of $G$ (c.f., \cite{Stanley:2012}, Chapter 9; \cite{vanLint:2001}, Chapter 36).  Since it illustrates the enumerative nature of the arguments used in the body of the text, we provide a proof here:
\begin{thm}[Matrix-Tree Theorem]
Let $\kappa(G,v)$ denote the number of oriented spanning trees of $G$ rooted at $v\in\mathcal{V}$, and $\Delta_{v}$ be the Laplacian matrix of $G$ with the $v^{\mathrm{th}}$ row and column deleted.  Then $\kappa(G,v)=\det\Delta_{v}$.
\end{thm}
\proof  Without loss of generality, let $v=n\in\mathcal{V}$.  Let the $(n-1)\times n$ matrix $F$ be the incidence matrix $E$ of $G$ with its $n^{\mathrm{th}}$ row deleted, so that $\Delta_{n}=FF^{\mathrm{T}}$.  Now, the Cauchy-Binet theorem (c.f., \cite{Stanley:2012}, Chapter 9) tells us that we can write:
\begin{equation*}
\det \Delta_{n}=\sum_{S}\det(F_{S})\det(F_{S}^{\mathrm{T}})=\sum_{S}\det(F_{S})^{2},
\end{equation*}
where $S$ ranges over all subsets of $\mathcal{E}$ of size $n-1$ none of which are outgoing from $n\in\mathcal{V}$, and $F_{S}$ is the $(n-1)\times (n-1)$ matrix whose columns are those of $F$ with indices in $S$.  This means that each choice of $S$ corresponds of a sub-graph $T\subset G$ with $n-1$ edges which have $d^{\mathrm{out}}_{n}=0$.  We now quote the following lemma from graph theory:
\begin{lemma}[c.f., \cite{Stanley:2012} Lemma 9.7]
Let $S$ be a set of $n-1$ directed edges of $G=(\mathcal{V},\mathcal{E})$, none of which are outgoing from $n\in\mathcal{V}$.  Then $\det F_{S}=\pm 1$ if $S$ forms a spanning tree of $G$ rooted at $n$, and $\det F_{S}=0$ otherwise.
\end{lemma}
This means that
\begin{equation*}
\det\Delta_{n}=\sum_{\mbox{\small{spanning trees rooted at n}}}(\pm 1)^{2}=\kappa(G,n),
\end{equation*}
as required.     $\Box$

The weighted generalizations of the Matrix-Tree theorem include weighted edges (e.g., \cite{vanLint:2001}, Chapter 36) and allow us to introduce the propagator appropriate to each edge.  This leads to the weighted Laplacians used throughout the paper.  We  define the weighted Laplacian matrix $\widehat{\Delta}$ for the directed graph $G$ with weights for each edge given by:
\begin{equation*}
\widehat{\Delta}=E\mathcal{W}E^{\mathrm{T}}, \mathcal{W}=\diag(w_{ij}), \widehat{\Delta}_{ij}=\left\{
\begin{array}{c}
-w_{ij} \;\mathrm{if}\;i\neq j\;\mathrm{and}\;(i,j)\in\mathcal{E}\\
\sum_{(i,k)\in\mathcal{E}}w_{ik}\;\mathrm{if}\;i=j \\
0 \;\mathrm{otherwise}
\end{array}\right. .
\end{equation*}
Then the weighted Matrix-Tree theorem reads:
\begin{thm}[Weighted Matrix-Tree Theorem]
Let $\mathcal{T}(G,v)$ be the set of all oriented spanning trees of $G$ rooted at $v\in\mathcal{V}$ and $\widehat{\Delta}$ be the weighted Laplacian of $G$.  For any $T\in\mathcal{T}(G,v)$, let $\mathcal{E}_{T}$ denote its set of directed edges.  Then
\begin{equation*}
\det\widehat{\Delta}_{v}=\sum_{T\in\mathcal{T}(G,v)}\left(\prod_{(i,j)\in\mathcal{E}_{T}}w_{ij}\right).
\end{equation*}
\end{thm}
When the weights are set to be twistor-string propagators, this is the primary result used in the text.


\section{Proof of Lemma \ref{xilemma}}
\label{Proof}

In this appendix we provide the proof of lemma \ref{xilemma} from the text:
\begin{lemma}
The quantity
\begin{equation*}
\sum_{i=1}^{n-2}\sum_{j=1,\;j\neq i}^{n-2}\mathcal{F}^{n-1\;n}_{i\;j}(\xi)\;\delta^{8|8}\left(\sum_{i=1}^{n}W_{i}\sigma_{i}\right)+ (n-1\leftrightarrow n),
\end{equation*}
is independent of choice of $\xi\in\CP^{1}$, where
\begin{equation*}
\mathcal{F}^{n-1\;n}_{i\;j}(\xi)=\frac{(\xi n)^{2}}{(n-1\;n)(in)}\left(\frac{(n-1\xi)}{(\xi i)(\xi j)^2}+\frac{(n-1\;j)}{(ji)(\xi j)^2}\right)\tilde{\Phi}_{ji}\;\mathrm{det}\left(\tilde{\Phi}_{n-1\;n\;i\;j}\right).
\end{equation*}
\end{lemma}

\proof The momentum conserving delta function ensures that the matrix $\tilde{\Phi}$ is independent of $\xi$, so we need only focus on the explicit $\xi$ dependence appearing in the definition of $\mathcal{F}^{n-1\;n}_{i\;j}(\xi)$.  The double sum over $i\neq j$ and $n-1\leftrightarrow n$ indicates that (term-by-term) we must symmetrize over $(i,j)$ and $(n-1,n)$.  Two applications of the Schouten identity leaves us with:
\begin{multline*}
\mathcal{F}^{(n-1\;n)}_{(i\;j)}(\xi)=-\frac{(\xi n)^{2}}{(n-1\;n)}\left(\frac{3(n-1\;\xi)(jn)(\xi i)+2(n-1\;\xi)(\xi j)(in)+(n-1\;i)(\xi j)(\xi n)}{(in)(jn)(\xi i)^{2}(\xi j)^{2}}\right) \\
\times \tilde{\Phi}_{ji}\;\mathrm{det}\left(\tilde{\Phi}_{n-1\;n\;i\;j}\right)+(n-1\leftrightarrow n).
\end{multline*}
This expression appears to have poles of order two at $\xi=\sigma_{i,j}$.  The residue is given by
\begin{multline*}
\lim_{\xi\rightarrow\sigma_{i}}\frac{\partial}{\partial\xi_{A}} \mathcal{F}^{(n-1\;n)}_{(i\;j)}(\xi) \\
=\left[-3(in)^{2}(n-1\;i)(j\;n-1)\sigma_{j}^{A}-2(in)^{2}(ij)(j\;n-1)\sigma_{n-1}^{A} \right. \\
+(n-1\;i)(in)(ij)(j\;n-1)\sigma_{n}^{A}+3(i\;n-1)^{2}(ni)(jn)\sigma_{j}^{A}+2(i\;n-1)^{2}(ij)(jn)\sigma_{n}^{A} \\
\left. -(ni)(i\;n-1)(ij)(jn)\sigma_{n-1}^{A}\right]\times\frac{\tilde{\Phi}_{ji}\;\mathrm{det}\left(\tilde{\Phi}_{n-1\;n\;i\;j}\right)}{(n-1\;n)(jn)(j\;n-1)(ij)^{2}}.
\end{multline*} 

Seven applications of the Schouten identity eventually reduce this expression to
\begin{equation*}
\frac{(i\;n-1)(ni)\sigma_{j}^{A}-(n-1\;i)(nj)\sigma_{i}^{A}-(i\;n-1)(ij)\sigma_{n}^{A}}{(jn)(j\;n-1)(ij)}\tilde{\Phi}_{ji}\;\mathrm{det}\left(\tilde{\Phi}_{n-1\;n\;i\;j}\right)=0,
\end{equation*}
so we find that
\begin{equation*}
\mathrm{Res}_{\xi=\sigma_{i}}\left(\mathcal{F}^{(n-1\;n)}_{(i\;j)}(\xi)\right)=0.
\end{equation*}
Since these are the only potential poles in the expression, we find that (with the momentum conserving delta function) $\mathcal{F}^{(n-1\;n)}_{(i\;j)}(\xi)$ is actually holomorphic in $\xi$, and therefore independent of $\xi$ by Liouville's theorem.  This completes the proof.     $\Box$


\section{Cachazo-Skinner Formula for $\cN=4$ Supergravity}
\label{N=4}

In this appendix, we illustrate how to reduce the Cachazo-Skinner formula for the tree-level S-matrix of $\cN=8$ supergravity to $\cN=4$ supergravity.  Recall the initial formula reads:
\be{CSF*}
\cM_{n,k} = \int \frac{\prod_{r=0}^{k+1} \d^{4|8}U_{r}}{\mathrm{vol}\;\GL(2,\C)}\mathrm{det}^{\prime}(\tilde{\Phi}^{k})\mathrm{det}'(\Phi^k)\prod_{i=1}^{n}\D\sigma_{i}\;h_{i}\left(\lambda_{i}\tilde{\lambda}_{i}; Z(\sigma_{i})\right),
\ee
where $h_{i}\in H^{0,1}(\PT,\cO(2))$ are the twistorial wavefunctions of the external particles.  To reduce to $\cN=4$ supersymmetry, we need to split the single $\cN=8$ graviton multiplet into two $\pm 2$ multiplets.  In this paper, our convention has been that the external states labeled $1,\ldots, n-k-2$ correspond to positive helicity gravitons and $n-k-1,\ldots,n$ to negative helicity gravitons, respectively.  

The maps $Z^{I}:\CP^{1}\rightarrow\PT$ have fermionic coordinates $\chi^{a}(\sigma)$, where $a=1,\ldots, 8$ is the $\cN=8$ $R$-symmetry index.  Let us write $\chi^{a}=(\chi^{\hat{a}}, \psi^{\mu})$, where $\hat{a}=1,\ldots, 4$ and $\mu=5,\ldots, 8$.  Then we can capture the appropriate splitting of the $\cN=8$ graviton multiplet into its $\cN=4$ subsectors on twistor space by taking
\begin{equation*}
\frac{\partial h_{i}}{\partial \psi^{\mu}} =0, \qquad i=1,\ldots,n-k-2,
\end{equation*}
and
\begin{equation*}
h_{i}=\psi^{4}\;\tilde{h}_{i}, \qquad \tilde{h}_{i}\in H^{0,1}(\PT,\cO(-2)), \qquad i=n-k-1,\ldots, n.
\end{equation*}
We also denote the $\mu$ portion of the moduli coordinates $U_{r}$ as $\nu^{\mu}_{r}$.  In this case, we have the following useful algebraic relation:
\be{fermints}
\int \prod_{r=0}^{k+1}\d^{4k+8}\nu_{r}\; \prod_{i=n-k-1}^{n}\psi^{4}_{i}\;\tilde{h}_{i}=\mathrm{VM}(n-k-1,\ldots,n)^{4},
\ee
where the right-hand side is the Vandermonde determinant defined by:
\be{VdM}
\mathrm{VM}(i,\ldots,j)\equiv \prod_{l<m,\; l,m\in\{i,\ldots,j\}}(lm).
\ee

Now, let us recall the definitions of the reduced determinants appearing in \eqref{CSF}:
\be{red-det}
\mathrm{det}'\left(\tilde{\Phi}^{k}\right)=\frac{\mathrm{det}\left(\tilde{\Phi}^{k}_{1\;n-k-1\cdots n}\right)}{\mathrm{VM}(1,n-k-1,\ldots,n)^2}, \qquad \mathrm{det}'\left(\Phi^{k}\right)=\frac{\mathrm{det}\left(\Phi^{k}_{1\cdots n-k-2\;n}\right)}{\mathrm{VM}(n-k-1,\ldots,n-1)^2},
\ee
where we have made convenient choices for the reduced determinants (other, equivalent, choices can of course be made \cite{Cachazo:2012pz}).  Combining \eqref{fermints} with these reduced determinants gives:
\begin{multline}\label{CSFN=4}
\cM_{n,k}=\int \frac{\prod_{r=0}^{k+1}\d^{4|4}U_{r}}{\mathrm{vol}\;\GL(2,\C)} \mathrm{det}\left(\tilde{\Phi}^{k}_{1\;n-k-1\cdots n}\right)\;\mathrm{det}\left(\Phi^{k}_{1\cdots n-k-2\;n}\right)\frac{\prod_{l=n-k-1}^{n-1}(ln)^2}{\prod_{m=n-k-1}^{n}(1m)^2} \\
\times \prod_{i=1}^{n-k-2}h_{i}\;\D\sigma_{i} \prod_{j=n-k-1}^{n}\tilde{h}_{j}\;\D\sigma_{j}.
\end{multline}
This is the $\cN=4$ truncation of the Cachazo-Skinner formula, and can easily be seen to coincide with the $k=0$ case explicitly derived in section \ref{MHV}.

\bibliographystyle{JHEP}
\bibliography{deSitter2}

\end{document}